\documentclass[preprint,10pt]{elsarticle}

\usepackage{amssymb}
\usepackage{amsfonts}
\usepackage{amsmath}
\usepackage{booktabs}
\usepackage{txfonts}
\usepackage{mathbbol}
\usepackage{mathrsfs}
\usepackage{epsfig,bm,dcolumn}
\usepackage{graphicx}
\usepackage{geometry}
\usepackage{color}
\usepackage{overpic}
\usepackage{slashed}
\usepackage[bookmarks,bookmarksnumbered]{hyperref}
\hypersetup{colorlinks = true,linkcolor = blue,anchorcolor =red,citecolor = blue,filecolor = red,urlcolor = red,
            pdfauthor=author}
\usepackage{subfig}
\usepackage[capitalise]{cleveref}

\usepackage{floatrow}
\usepackage{subfig}
\journal{XXX}

\begin{document}

\begin{frontmatter}



\title{Machine learning spatio-temporal epidemiological model to evaluate Germany-county-level COVID-19 risk}
\author[fias]{Lingxiao Wang}
\author[buaa1]{Tian Xu}
\author[bh]{Till Hannes Stoecker}
\author[fias,itp]{Horst Stoecker}
\author[buaa1,buaa2]{Yin Jiang\corref{cor}}
\author[fias]{Kai Zhou\corref{cor}}

\cortext[cor]{Corresponding author.}

\address[fias]{Frankfurt Institute for Advanced Studies, Ruth-Moufang-Str. 1, 60438 Frankfurt am Main, Germany}
\address[buaa1]{Department of Physics, Beihang University, Beijing, 100191, China}
\address[bh]{Black Hole KG, 61440 Oberursel (Taunus), Germany}
\address[itp]{Institute for Theoretical Physics, Goethe University Frankfurt, 60438 Frankfurt am Main, Germany}
\address[buaa2]{Beijing Advanced Innovation Center for Big Data-Based Precision Medicine, School of Medicine and Engineering, Beihang University, 100191, China}

\begin{abstract}
As the COVID-19 pandemic continues to ravage the world, it is of critical significance to provide a timely risk prediction of the COVID-19 in multi-level. To implement it and evaluate the public health policies, we develop a framework with machine learning assisted to extract epidemic dynamics from the infection data, in which contains a county-level spatiotemporal epidemiological model that combines a spatial Cellular Automaton (CA) with a temporal Susceptible-Undiagnosed-Infected-Removed (SUIR) model. Compared with the existing time risk prediction models, the proposed CA-SUIR model shows the multi-level risk of the county to the government and coronavirus transmission patterns under different policies. This new toolbox is first utilized to the projection of the multi-level COVID-19 prevalence over 412 Landkreis (counties) in Germany, including t-day-ahead risk forecast and the risk assessment to the travel restriction policy. As a practical illustration, we predict the situation at Christmas where the worst fatalities are 34.5 thousand, effective policies could contain it to below 21 thousand. Such intervenable evaluation system could help decide on economic restarting and public health policies making in pandemic.


\end{abstract}


\end{frontmatter}

\section{Introduction}\label{sec:1}
Public health security is the cornerstone of economic and social stability. CoronaVirus Disease 2019 (COVID-19), an infectious disease caused by the Severe Acute Respiratory Syndrome Coronavirus 2 (SARS-CoV-2), became a global pandemic that spread out across the world from the beginning of 2020. Since middle of March, the number of infected cases in Germany experienced a rapid growth. Although the severe restriction of travel eases the situation in the summer, the number of infected cases have been increasing again with a more critical trend since mid-September, named routinely as the second-wave pandemic worldwide. By November 25, 2020, the pandemic has caused  983588 confirmed cases and 15160 fatalities in Germany. Many states are facing dilemma challenges of public heath and economics in curbing the spread of the COVID-19, e.g., Berlin, Bavaria and Hessen. The medical resources of Germany (ICU beds and doctors) have not reached their ceiling, but the current phase of the pandemic is not encouraging. Being a high-infectious disease, COVID-19 is expected to continue spreading or be even worse in the coming winter, which will cause a higher number of infections and mortality in the next few months. With no perfect medical treatments nor enough vaccines available until now, implementation of relatively effective public health policies such as social distancing and travel restriction become the primary measures of the states to mitigate the spread of epidemic, which has been proven to be an effective containment in East Asia, e.g. China~\cite{maier:2020effective} and South Korea, also in Europe at early stage~\cite{walensky:2020mitigation}. \textbf{In this worsening health crisis, a dire need emerges to establish a multi-level (Nation-State-County) COVID-19 risk prediction model, which can guide the evaluation and improvement on the public health policies of governments (states) at multi-level.} Such intervenable risk evaluation model is beneficial to policymakers in decision-making on economics restarting and public health policies at different scales during epidemics. The multi-level risk information can help governments to improve the public restriction strategies such as cross-state travel restriction, home office rules, severe home isolating regulations and vaccination resources allocation. Residents, who are concerning the present situation, can also benefit from the multi-level risk information, to make effective access of medical resources and help implementing or following the public health policies.

Evaluation to the risk of the on-going COVID-19 pandemic across the geographical-dependent administrative divisions is challenging because of the substantial heterogeneity rooted in economic composition, industrial diversity, traffic conditions and political views. Nonetheless, it is inevitably an emergent demand for governments at all levels. Although a lot of studies have been devoted to the dynamics of the epidemics based on mathematical models~\cite{barbarossa:2020modeling,barbarossa:2020first},  (see also reviews\footnote{Since the number of COVID-19 papers are too many to list them all, the following references are highly relevant to the topic we discussed in this paper.}~\cite{estrada:2020covid19,vespignani:2020modelling,wang:2020review}), they mainly focus on time-domain analysis and ignore the influence of geographic distribution and population migration~\cite{walensky:2020mitigation} which palpably presents as cross-county transfers at daily-scale.\textbf{ A model which can predict the spatio-temporal evolution of the COVID-19 cases is urgently needed to assess the local risk  at multi-level regions.} Besides works which concern spatial distribution or local evolution based on the historical statistical data~\cite{kergassner:2020mesoscale,linden:2020foreshadow,long:2020multifractal} , recently there are several studies sharing similar goals in different countries~\cite{chande:2020realtime,rader:2020crowding,wang:2020global,zhou:2020spatiotemporal}, which is merited to be one of the most urgent needs across the world. In a practical perspective, a reasonable and efficient risk prediction model needs to be performed comprehensively with all Landkreis~\footnote{It is the primary administrative subdivision higher than a city in Germany, also could be named as county in English.}, since people is highly mobile and connected by highways, railways and airways in Germany. Conceivably, a county-level risk evaluation to the public and local governments is informative and ponderable. 

Machine learning (ML), a branch of artificial intelligence, efficiently integrates statistical and inference algorithms and thus offer the opportunity to uncover hidden structure of evolution in complex coronavirus data and to describe it with finite dynamical parameters. Therefore, combinatory usage of ML algorithms and spatio-temporal epidemiological model for risk prediction in the context of COVID-19 pandemic are worth exploring. Although the state-of-the-art ML techniques have been deployed into myriad fields of COVID-19 pandemic~\cite{gao:2020machine,shen:2020using,yesilkanat:2020spatiotemporal,zou:2020epidemic}, for the governments and residents, the highest concern is how fast the COVID-19 diffuses. Available ML models that focus on this exhibit promising implications in individual or national scale, but are still impeded by the paucity of validation information and limited privacy rules in Germany, thus lack the capability of predicting multi-scale evolution.

\textbf{In this study, we aim to establish a new evaluation paradigm by introducing a spatio-temporal epidemiological model with machine learning assisted.} The dynamical model of the epidemic contains a Cellular Automaton (CA)~\cite{white:2007modeling,schneckenreither:2008modelling} and a modified Susceptible-Undiagnosed-Infected-Removed (SUIR) model~\cite{wang:2020global,jun-feng:2020assessment}. The SUIR model is more inclusive than the previous Susceptible-Infectious-Recovery(SIR)-type models since consideration of the Undetected or Asymptomatic populations~\cite{chen:2020timedependent,ivorra:2020mathematical} hidden under the lack of detection. As an important extension of the SIR model, we divide the infectious population into two compartments: one is the undiagnosed who manifest as normal susceptible individuals but with possible infectious and no timely detection; the other one is the infected (diagnosed) who are confirmed by detection and have been admitted to hospital or isolated at home. In this new temporal model, we constrain the mobility of the people in the susceptible and undiagnosed populations, in which the understanding to the latter will help address the underestimation issue concerning infection cases in the public databases. As a matter of fact, the movement of the undiagnosed people across the counties will contribute to the deteriorating epidemic situation, thus we introduce the CA model to capture the influences of the population mobility. Being assisted with ML to extract the dynamical parameters, this new evaluation system can inform the public and governments in different regions with effective warning, which should be considered into adjusting the restriction rules. 

The article is organized as follows. In Section.~\ref{sec:2}, we propose a full framework of risk prediction to COVID-19 pandemic in Germnay with introducing the spatio-temporal CA-SUIR model. The machine learning techniques are applied to extract the dynamical parameters from reported data of Robert Koch-Institut (RKI). Section.~\ref{sec:3} reproduces the evolution of infection maps from 25 February to 28 March 2020, and extracts the infectious parameters from RKI data in September 2020 with a well-trained machine which was training on the data set from CA-SUIR model simulations. The well-learned machine predicts county-level infection maps from September to November 2020, which is matched remarkably with the reported situation. With regard to the up-coming Christmas holiday, the different public health policies are evaluated in the end of the section, in which the entire, partial and none lock-down strategies are compared. In the last section, some concluding remarks are presented.

\section{Machine Learning Spatio-temporal Epidemiological Model}\label{sec:2}

\begin{figure}[htb!]
	\centering
	\includegraphics[width=15cm]{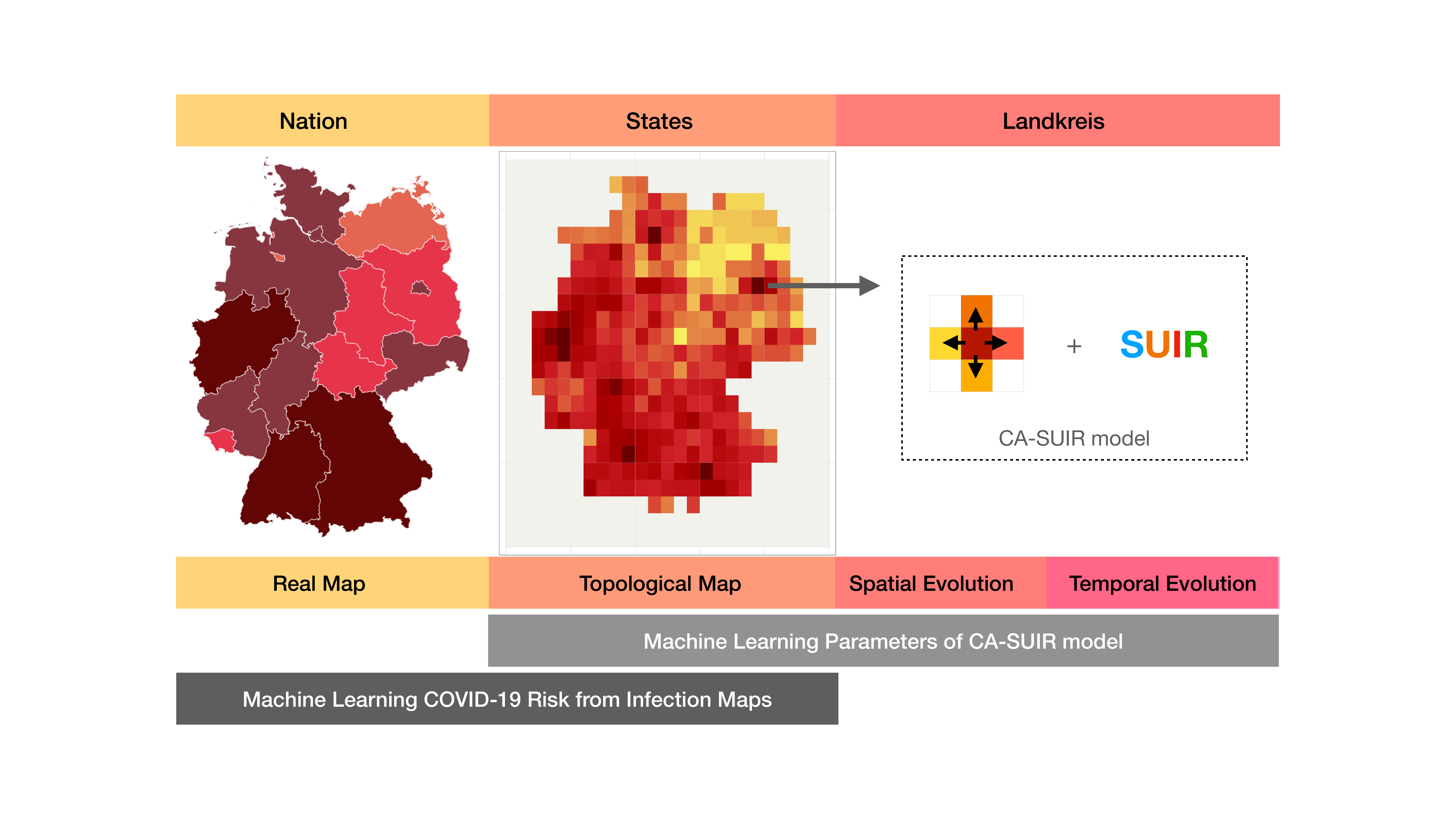}
	\caption{\textbf{A framework to evaluate and predict the COVID-19 risk in Germany.} The left panel is the geographic distribution of the infection cases in Germany, which is named as the real map at nation level. The central panel is the topological equivalent map of the infection cases, in which the main cities are mapping onto the sites on square lattice at state level. The right panel is the microscopic dynamics on each site, which is spatially depicted by Cellular Automata and temporally evolved through SUIR model at county level. The color of the characters in SUIR corresponds to the subsequent Fig.~\ref{fig:SUIR}.}
	\label{fig:flowchart}
\end{figure}
In this section, we propose a spatio-temporal model to predict the multi-level COVID-19 risk in Germany, named as CA-SUIR model in the following parts. Machine learning techniques are introduced to learn the dynamical parameters of the CA-SUIR model given a 'movie' of infection maps, which is sequentially utilized to evaluate the COVID-19 risk from RKI infection map timely in Germany. The full flow chart is assembled in Fig.\ref{fig:flowchart} to present the multi-level spatio-temporal simulation of epidemics, where as the first step real maps are projected onto the topological equivalent maps as explained in Appendix.~\ref{ref:top}. The cumulative infected cases in the real profiles\footnote{see the map at \href{https://en.wikipedia.org/wiki/COVID-19_pandemic_in_Germany}{COVID-19 pandemic in Germany}} are reassembled into a topological equivalent map with $25\times25$ sites in Germany, which contains 412 counties (see details in Appendix.~\ref{ref:top}). The micro-states of each county, its number of infected cases, are driven by the SUIR and CA simultaneously in the model, which construct the topological infection map. With simulated data from the CA-SUIR model, the Conv2LSTM neural networks~\cite{shi:2015convolutional} consist of the Time-distributed Convolution Neural Networks(CNNs) and the Recurrent Neural Networks(RNNs) are designed to learn to extract dynamical parameters of CA-SUIR model. Subsequently, we transfer the well-trained neural networks to make timely prediction and risk evaluation on RKI infection maps.

\subsection{Deep learning dynamical parameters}

\begin{figure}[htbp!]
	\centering
	\includegraphics[width=16cm]{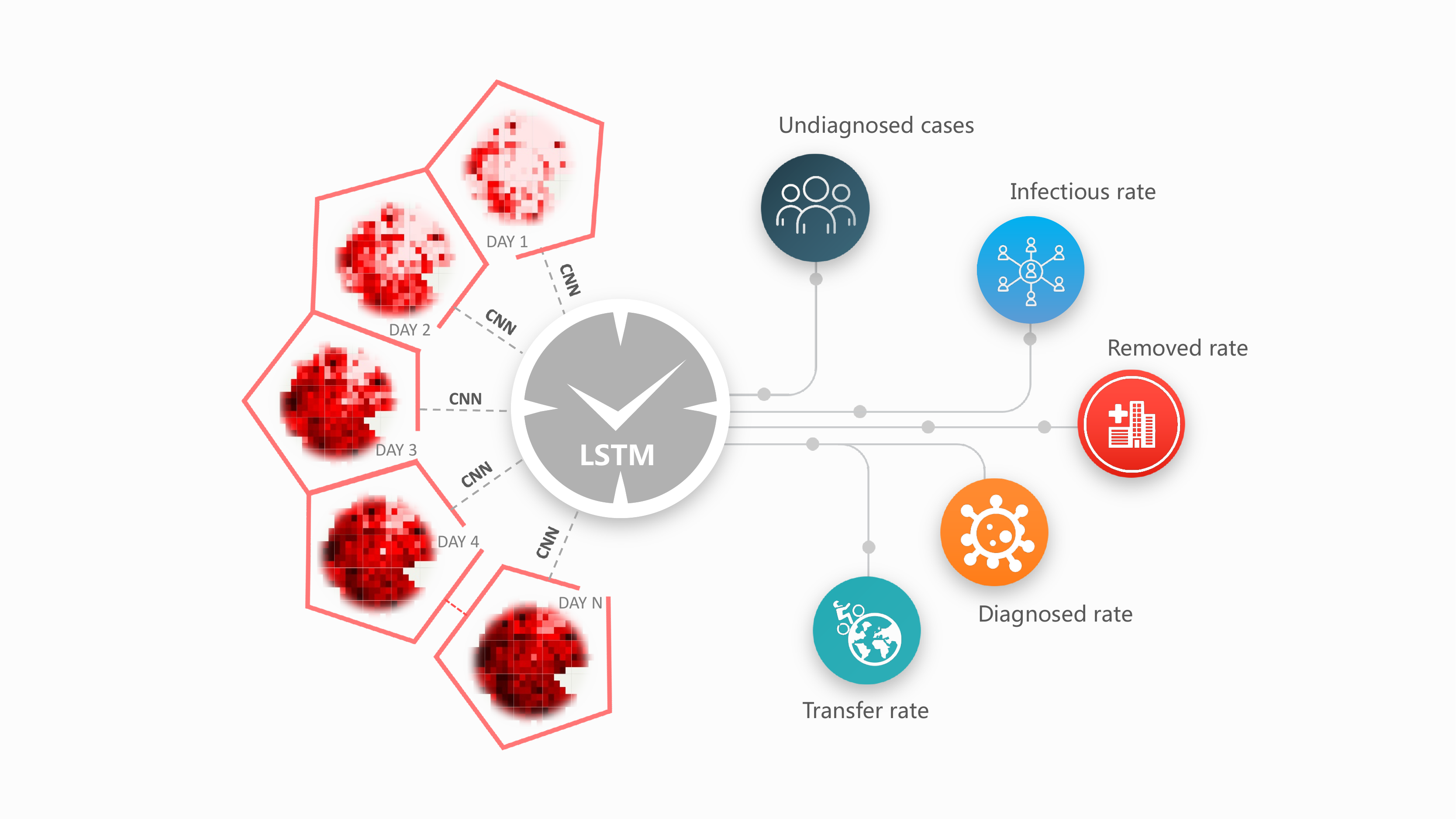}
	\caption{\textbf{Machine learning the dynamical parameters of CA-SUIR model from the evolution of infection maps}. The Time-Distributed Convolutional Neural Networks(CNNs) layers and Long Short-Term Memory(LSTM) units are combined to extract the information encoded in the input short movie (one image per frame). The undiagnosed population, infectious rate, removed rate, diagnosed rate and transfer rate are decoded from the Conv2LSTM neural networks.}
	\label{fig:dl}
\end{figure}
CNN model and Long Short-Term Memory (LSTM) model have been popularly adopted and show remarkable performance in many researches like in computer vision or Natural Language Processing tasks. While the CNN model can do efficient pattern recognition with convolutional filters on image-like tensor data, the LSTM model is capable of capturing dependencies along time direction between sequential tallies. In this work we combine CNN and LSTM model into Conv2LSTM for analysing the infection dynamical evolution.

The Conv2LSTM neural networks we built are demonstrated in Fig.~\ref{fig:dl}, in which the Time Distributed Convolutional Neural Networks(CNNs) layers and LSTM layers are combined to decode the information encoded in the short movie constructed by the infection maps. The details of the structure are arranged in  Appendix.~\ref{ref:conv2lstm}. The time distributed CNNs could extract features (with temporal structure preserved) from sequential image inputs and then embed them into the following LSTM layers, which is able to efficiently drill evolution rules from the short movies.

\subsection{Temporal SUIR model}
\begin{figure}[htb!]
	\centering
	\includegraphics[width=7cm]{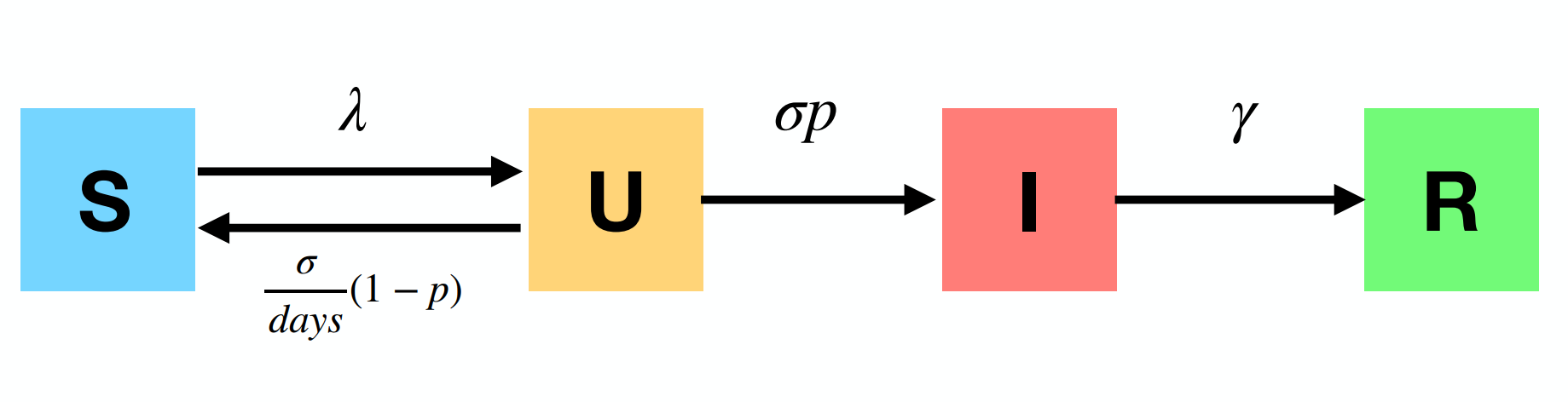}
	\caption{The SUIR diagram shows how individuals are infected and transferred among each compartments in a close area.}
	\label{fig:SUIR}
\end{figure}

We propose a proposed modified Susceptible-Undiagnosed-Infected-Removed (SUIR) model and demonstrate it in Fig~\ref{fig:SUIR}, where the infectious rate $\lambda$ controls the rate of spread which represents the probability of transmitting disease between a susceptible and an infectious individual. The diagnosed rate $\sigma$ dictate the probability of latent infectious individuals becoming confirmed by testing. Removed rate $\gamma$ contains the rate of mortality and recovery from the infectious.  Besides the classical definition of SIR model, considering the special features of COVID-19, we introduce the modified SUIR model: there is a considerable population that could be infected but without any symptoms, or, they just have no access to perform a detection(with possibility $p$), which is concluded in a new compartment, Undiagnosed(U). In contrast to the Infected who is usually subjected to be isolated in medical facilities or quarantine at home, the Undiagnosed population is with normal mobility, which leads to the high infection; the infectivity could decrease after $d$ days~\cite{hernandez-vargas:2020inhost}, but there is no clear evidence that they will get special immunity, which means immunity loss.

In a closed population with no births or mortalitys per day,the modified SUIR model becomes,
\begin{align}
\frac{d S}{d t} &=-\frac{S}{N}\lambda U + \frac{\sigma'}{\text{days}}U\\
\frac{d U}{d t} &=\frac{S}{N}\lambda U -\sigma U- \frac{\sigma'}{\text{days}}U  \\
\frac{d I}{d t} &=\sigma U-\gamma I \\
\frac{d R}{d t} &=\gamma I
\end{align}
where $N = S + U + I + R$ is the total population. The redefinition parameters in Fig.\ref{fig:SUIR} are $\sigma \equiv \sigma p, \sigma' \equiv \sigma(1-p)$. The detail explanation of the parameters are in Appendix.~\ref{ref:para}.

\subsection{Spatio-temporal CA-SUIR model}
\begin{figure}[htb!]
	\centering
	\includegraphics[width=8cm]{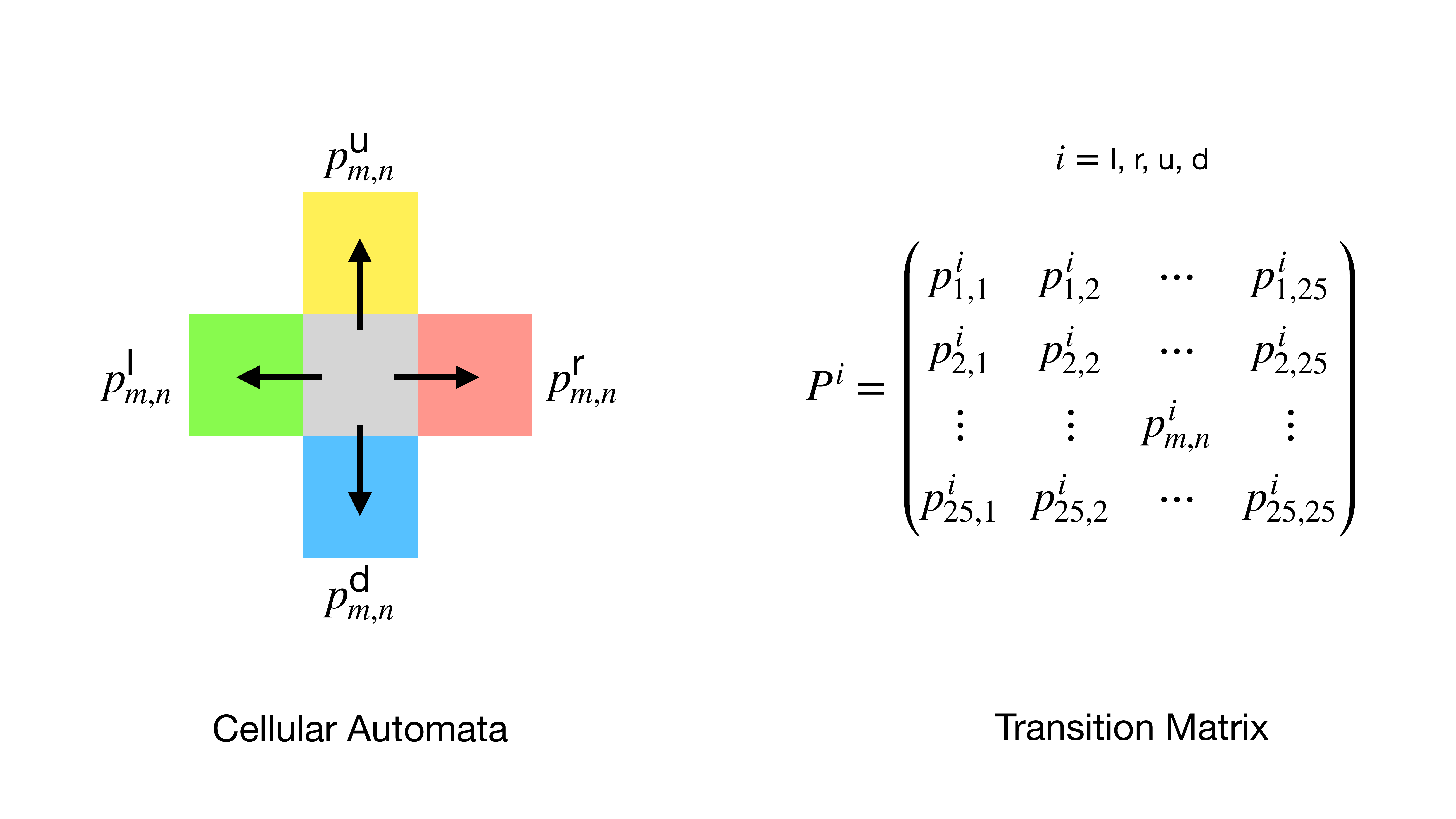}
	\caption{\textbf{Transition matrices in Cellular Automaton  model.} The size of the grid is $L=25$ in our case.}
	\label{fig:CA}
\end{figure}

We construct a cellular automaton model for tracking the epidemics spatial evolution in a closed system (Nation with fixed boundary as Fig.~\ref{fig:flowchart} shown, which matches the extreme restriction adopt by the German government from March 16. to May). The underlying structure is a $ L\times L $ cell grid, where $ L $ is the system size, and the number is from the topological map. The color of cell characterizes the population in minimum geographical unit, which is county in our case. It's an instructive example to set the cell size to a typical area of city, such as $10^2\sim10^3 \text{km}^2$, which can simulate the epidemics in the corresponding resolution approximately. The resolution could be naturally increased into community level, but it needs more detailed data which is laboriously accessible, thus the following discussion focuses on county level and corresponding resolution. This model adopts the Moore neighbor, and cells update their states by transition tensor $ P_{m,n}^i(t) $ with $ 4 \times L\times L $ shape, which means the possibility of transfer from city $ (m,n)$ to direction $i$ at $t$ time-step. In Fig.~\ref{fig:CA}, the $ P^i(t) $ shows as 4 matrices $(P^\text{u}(t),P^\text{d}(t),P^\text{l}(t),P^\text{r}(t))$ with same lattice size, in which different colors labels the direction of transfer. In the following simulation, we adopt the mean field approximation~\cite{wang:2019escape}, which leads to the uniform transition possibility $P^\text{u}(t) = P^\text{d}(t) = P^\text{l}(t) = P^\text{r}(t) = \epsilon$. The transfer rate $\epsilon$ is introduced to describe the average movement of the residents on each site. To mimic the border control induced closed boundary, a transfer rate $\epsilon = 0$ has been used in the primary epidemic stage~\cite{eckardt:2020covid19}.

The evolution of the CA-SUIR model is as follows,

\textbf{Step.1 Initialization.} Set the position of infection burst $ (x, y) $ and generate population distribution $ S(t) $ at the $ L\times L $ lattice. At $ t=0 $ time-step, epidemic turns on and population begins to migrate;

\textbf{Step.2 SUIR dynamics.} At $ t $ time-step, evolve SUIR model $t'$ times on each site;

\textbf{Step.3 Movement.} The parts of susceptible and undiagnosed people (S and U) on site $ (m,n) $ migrate into the neighbor site with transition matrices $ P^k(t,m,n) $ at $ t+1 $ time-step. Update all sites synchronously;

\textbf{Step.4 Update.} Turn to \textbf{Step.2} until $ T $ time-step is reached.

\section{Results}\label{sec:3}
\subsection{Pandemic dynamics modeling, data set generation and network capacity}

\begin{figure}[htb!]
	\centering
	\includegraphics[width=6.5cm]{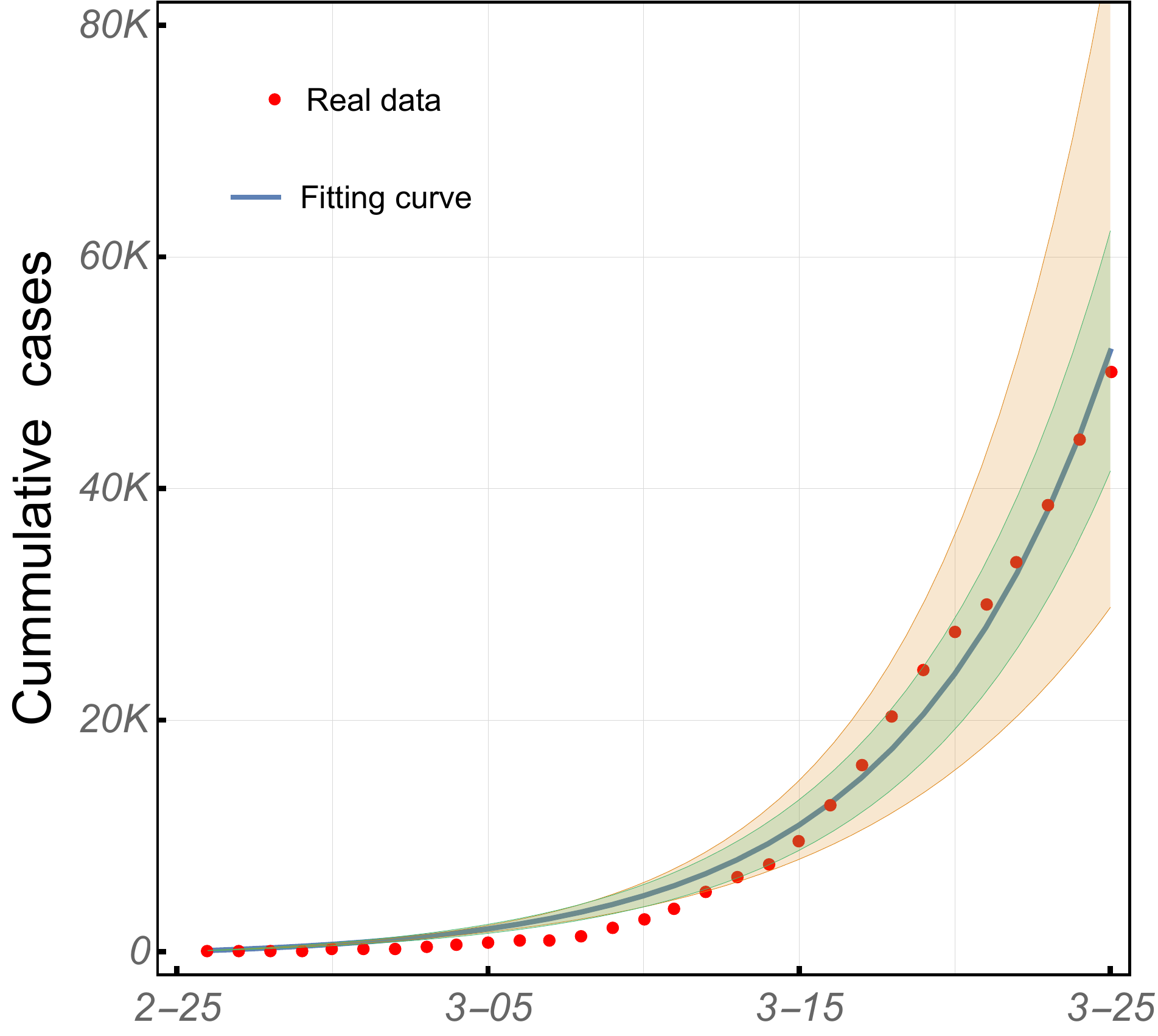}
	\caption{\textbf{The comparison of RKI data and simulated cumulative infection cases.} The data is from 25 Feb.2020 to 28 Mar.2020, in which the shadow bars are from uncertainty estimation. The orange panel is the variation space of SUIR model parameters, which shows a reliable area under $10\%$ changes. The green panel is from the $10\%$ uncertainty of the undiagnosed population(U), which is actually from the stability issues of RT-PCR testing~\cite{li:2020stability} and unreported cases in practical situation.}
	\label{fig:total}
\end{figure}

\begin{figure}[htbp!]
	\centering
	\sidesubfloat[]{\includegraphics[width=4.5cm]{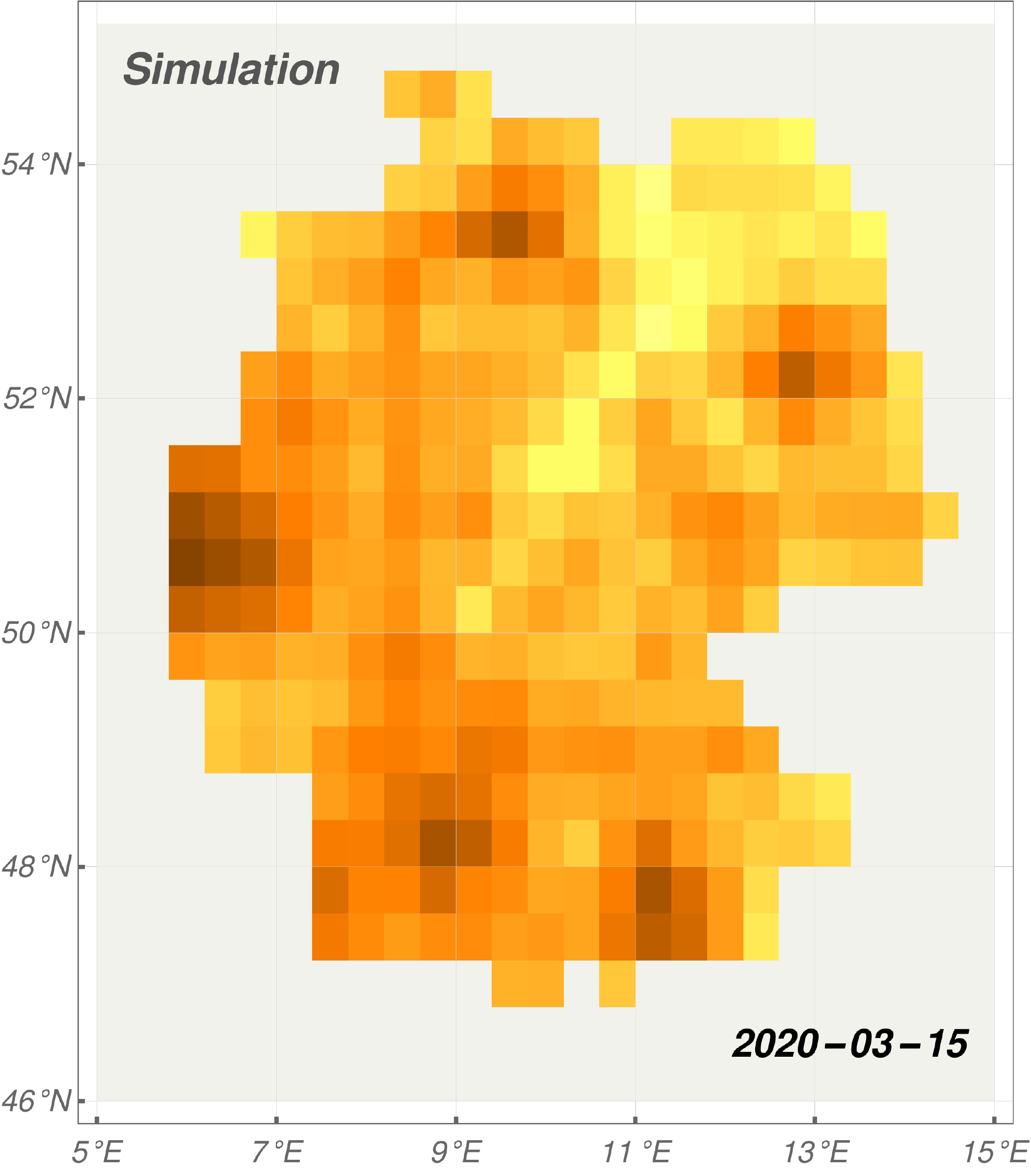}}
	\hspace{0.5 cm}
	\sidesubfloat[]{\includegraphics[width=4.5cm]{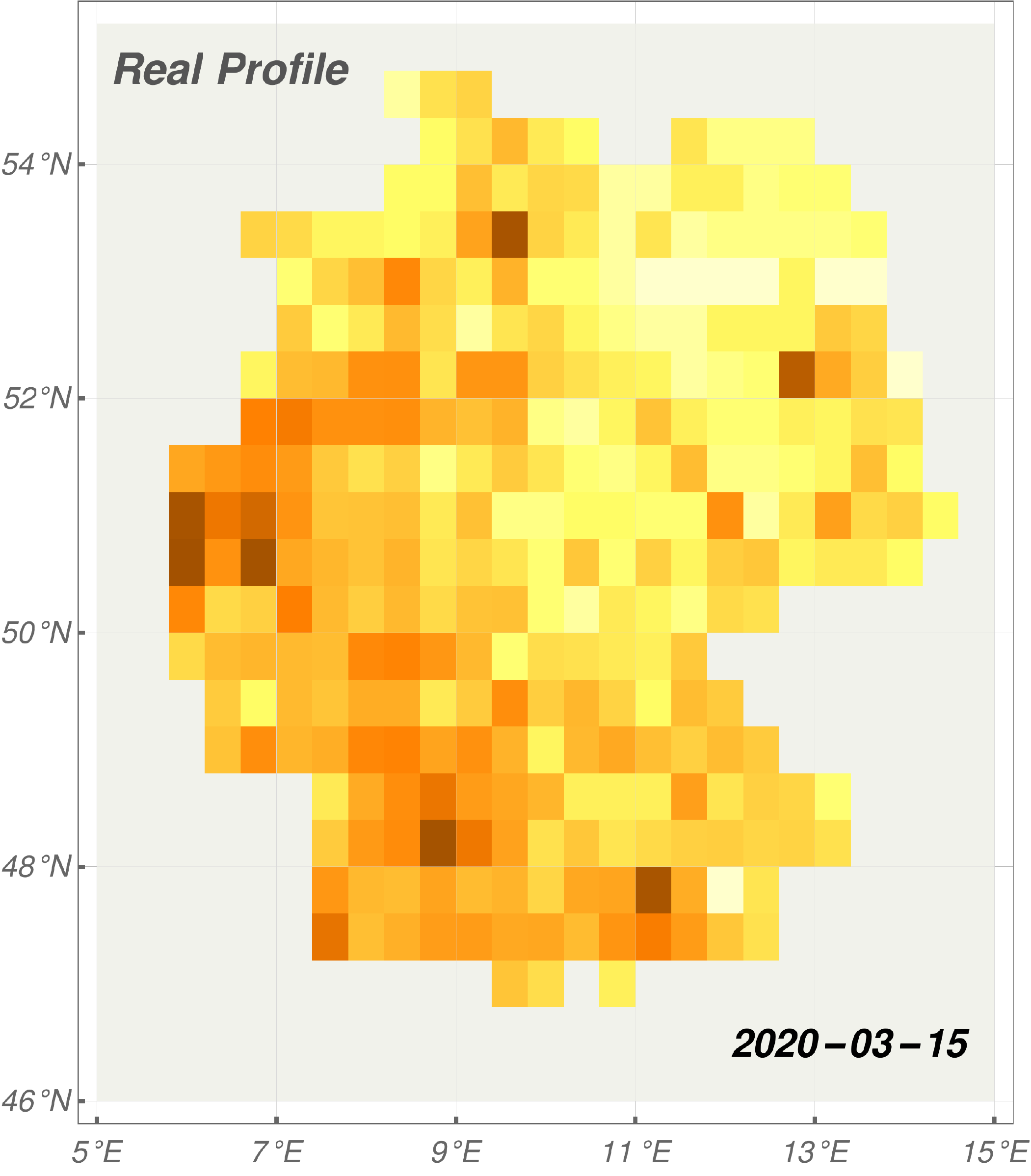}
	\hspace{0.25 cm}
	\includegraphics[scale = 0.3]{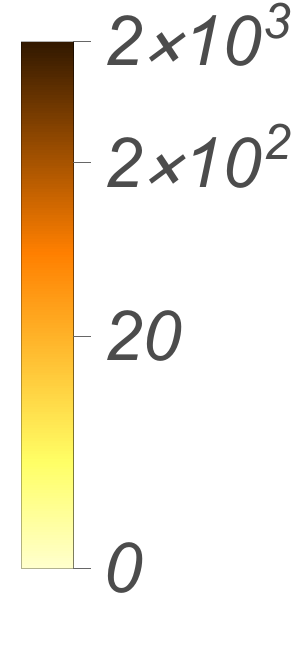}}
	\caption{\textbf{Simulated and data evolution of the infection distribution in Germany.} The infection maps are evaluated through CA-SUIR model, from the same initial distribution at 25 Feb.2020. The color bar is set in an uniform orange color scale, the darkest color is 2000 cases with a logarithmic re-scale. The horizontal and vertical coordinates correspond to the real geographical latitude and longitude, which is consistent in the following figures.}
	\label{fig:distro}
\end{figure}

With the daily number of cumulative cases in Germany (Data is from the open database of Robert Koch-Institut (RKI)~\footnote{ \url{www.rki.de/covid-19} and \href{https://covid-19-geohub-deutschland-esridech.hub.arcgis.com/datasets/9644cad183f042e79fb6ad00eadc4ecf_0/data}{COVID-19 GeoHub Deutschland}}), we fix the SUIR parameters to be $\lambda' \equiv \lambda - \frac{N}{S} \sigma'/\text{days} = 0.293, \sigma  = 0.129, \gamma = 0.628$ by fitting from redefined exponential behavior parameters(see  Appendix.~\ref{ref:para}). This parameter set can fit well the first-wave burst out in March especially the rapidly increasing stage (from 25 Feb.2020 to 28 Mar.2020), as presented in Fig.~\ref{fig:total}. The number of initial undiagnosed cases is determined in the model as $U_0\simeq 700$ which distributes as the infection cases. Fig.\ref{fig:total} suggests that there is an overestimation of the simulation at early stage(first 20 days), which might be naturally understood to be related with the underestimation in reality at that time since the limited detection ability and the supply of test reagents were insufficient~\cite{vandenberg:2020considerations}. Conceivably, by introducing more practical factors from clinical diagnosis or management, the SUIR model can be predictably improved to match with the record data, which will be discussed in our further works but is not the focus in the current one. According to the initial infection cases distribution and the locations of main cities of Germany, we set the initial undiagnosed population distribution $U_0(m,n)$ to be proportional to the population distribution of the cities $N(m,n)$. In the simulations for March we choose the transfer rate $\epsilon = 0.08$ as a constant, the population of transfer on site is proportional to the rate in each time-step. With the fitted parameters from above SUIR model, we generate the simulated evolution of the cumulative infection cases map in Fig.\ref{fig:distro}, which demonstrates the high consistency with the RKI data map (named as real profiles, see the full evolution sample in Appendix.~\ref{ref:top}).

\begin{figure}[htbp!]
	\centering
	\sidesubfloat[]{\includegraphics[width=5cm]{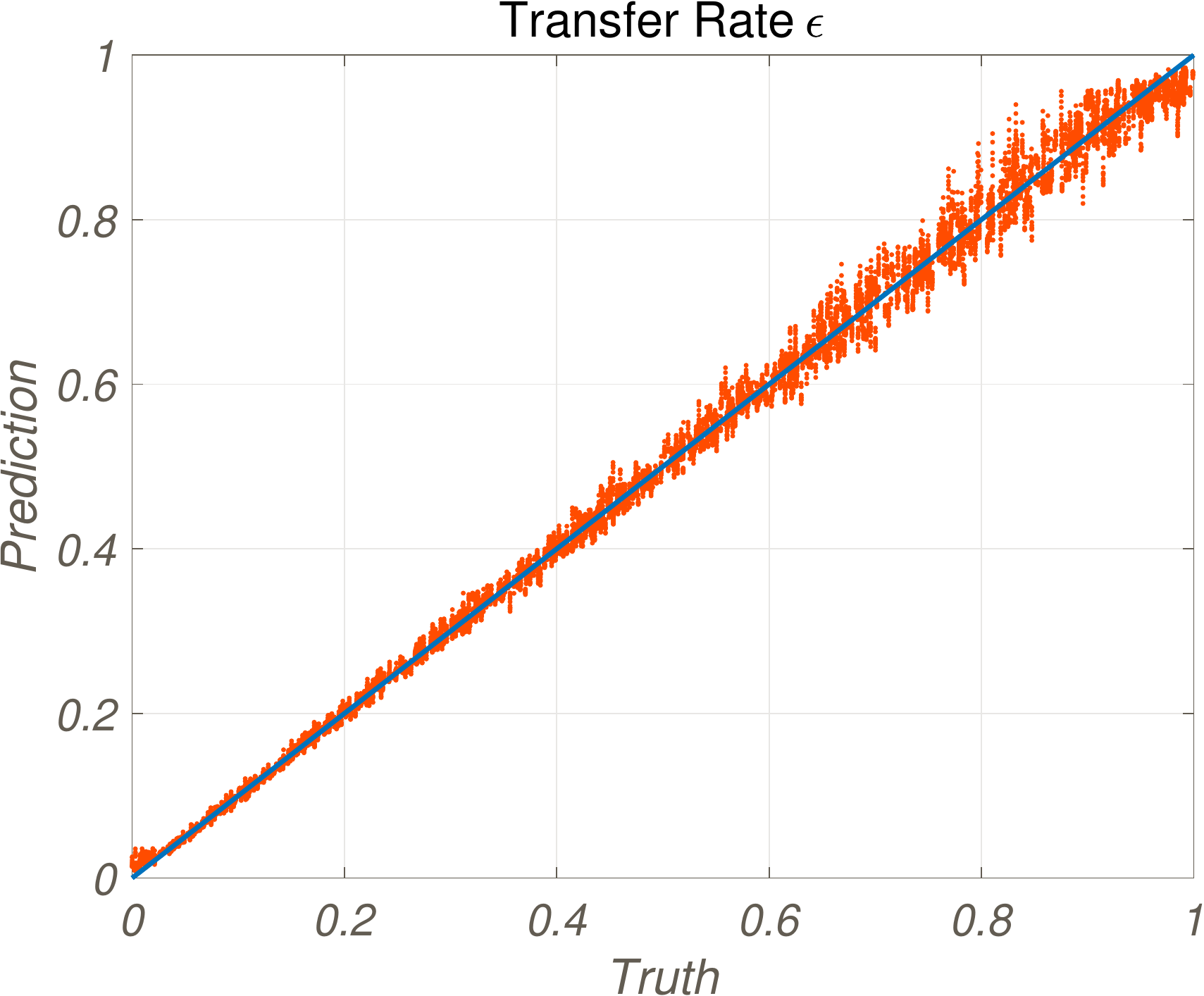}\label{fig:epsilon}}
	\hspace{0.1 cm}
	\sidesubfloat[]{\includegraphics[width=5cm]{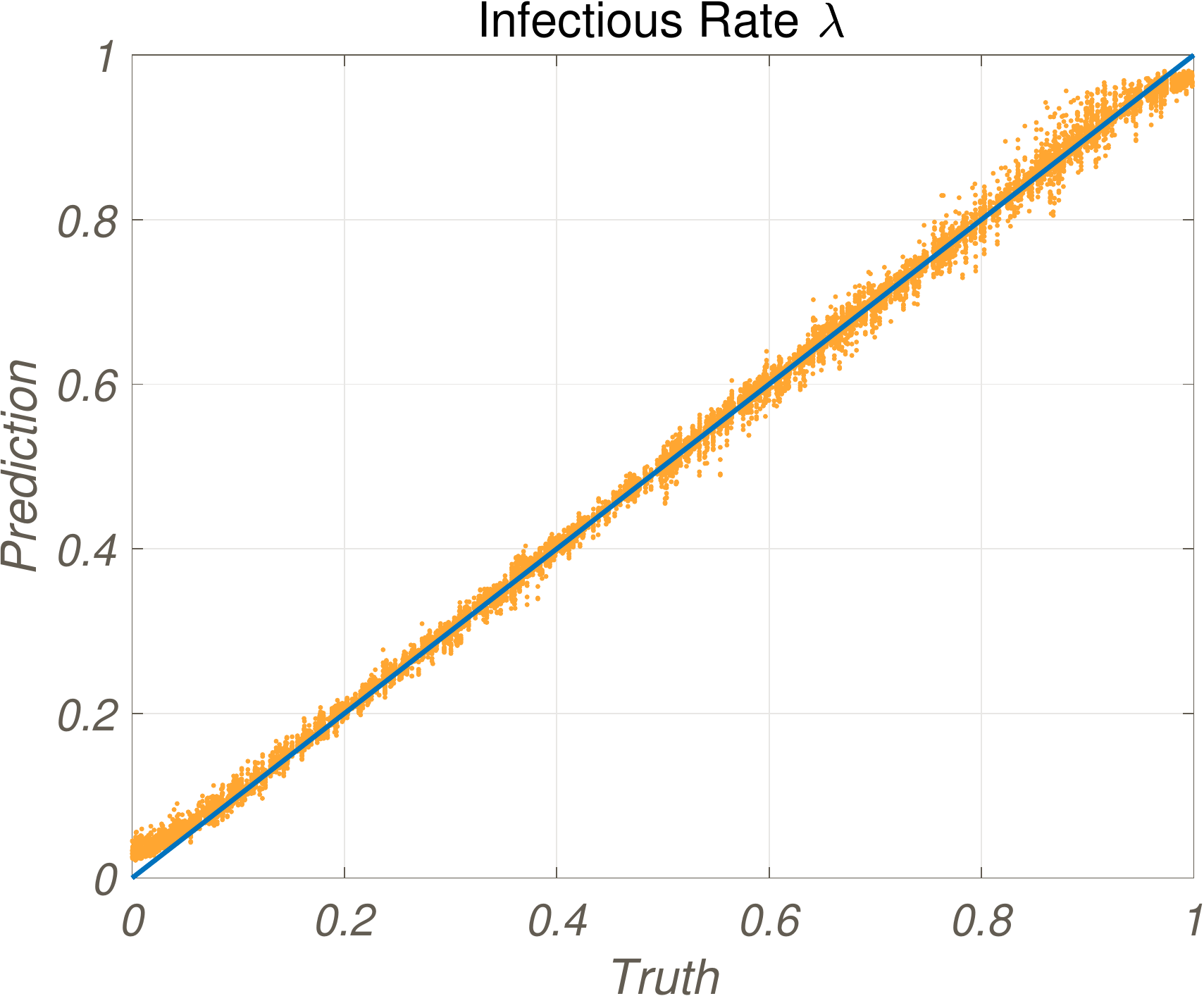}\label{fig:lambda}}
	\vspace{0.5 cm}
	\hspace{0.1 cm}
	\sidesubfloat[]{\includegraphics[width=5cm]{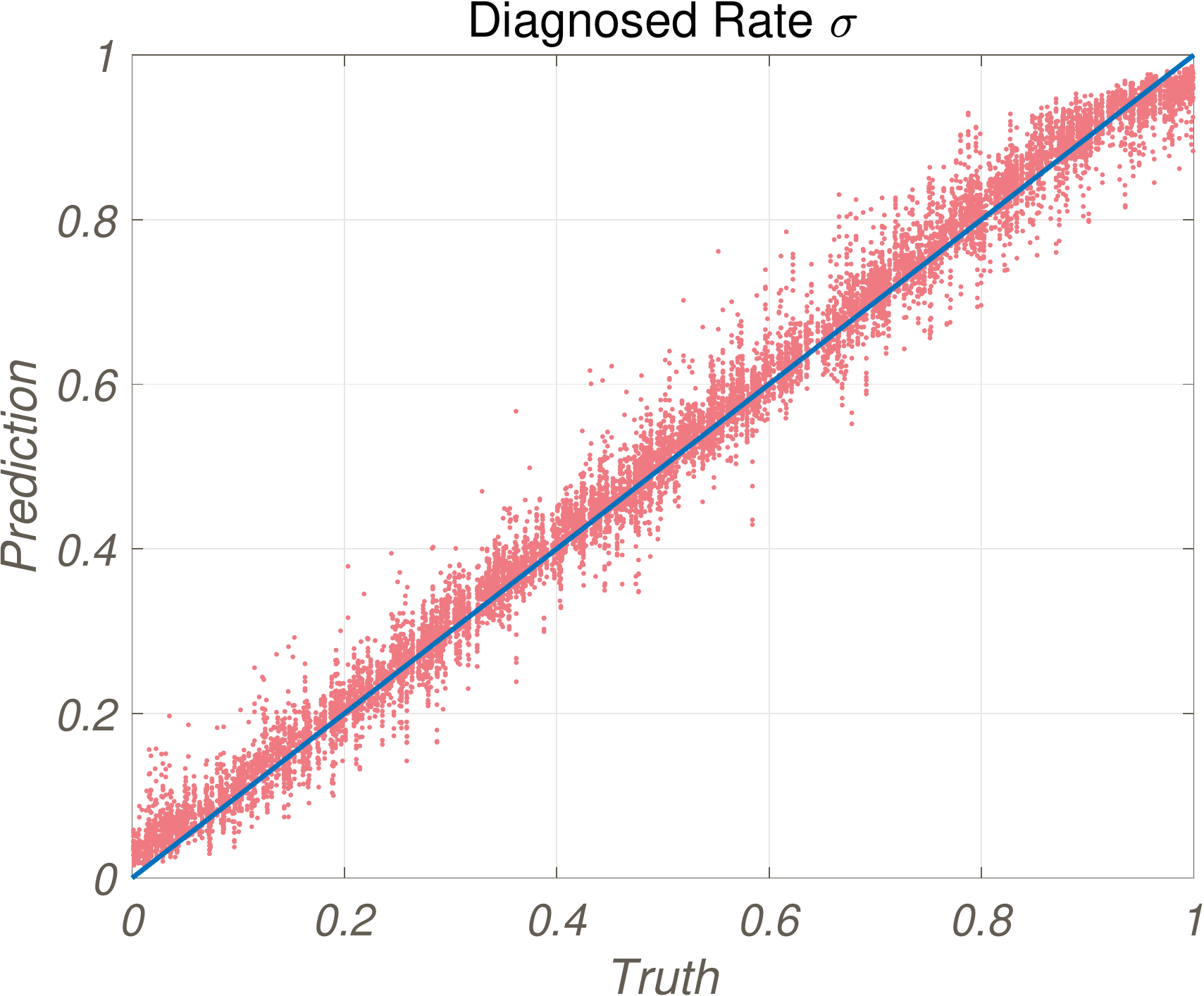}\label{fig:sigma}}
	\hspace{0.1 cm}
	\sidesubfloat[]{\includegraphics[width=5cm]{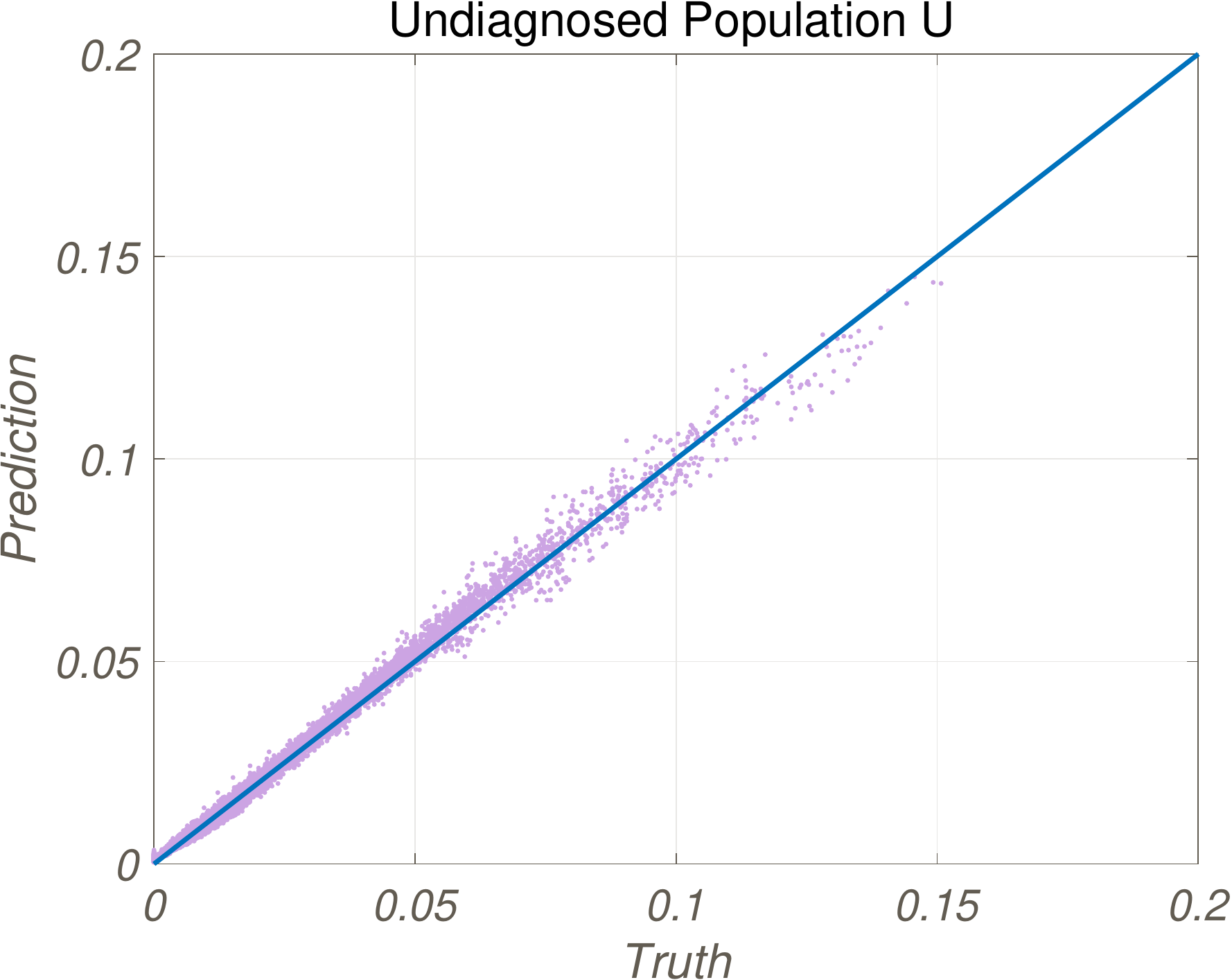}\label{fig:un}}
	\caption{\textbf{The performance of Conv2LSTM neural networks to extract dynamical parameters from infection maps.} \textbf{(a)} The transfer rate $\epsilon$ is tested on the testing data-set with the coefficient of determination $R^2=0.996$. \textbf{(b)} The infectious rate $\lambda$ is tested on the testing data-set with the coefficient of determination $R^2=0.998$. \textbf{(c)} The diagnosed rate $\sigma$ is tested on the testing data-set with the coefficient of determination $R^2=0.987$. \textbf{(d)} The undiagnosed population $U$ is tested on the testing data-set with the coefficient of determination $R^2=0.993$.}
	\label{fig:test}
\end{figure}

To prepare training data for Conv2LSTM, we generate 30-day simulation maps as short movies by associating the SUIR model parameters a $10\%$ uncertainty range. The total number of the data set is 10000, which contains 9000 infection movies as training data-set and the rest 1000 as testing data-set. The 30-day movie contains 30 frames, which is fed to the machine with continuous 7-day segments as slip windows~\cite{ramos:2017deepvel}. The structure of the Conv2LSTM neural networks are exhibited in Appendix.~\ref{ref:conv2lstm}. With the short movies to be the input, the Conv2LSTM is trained to predict the involved epidemiological model simulation parameters. The learning curve is shown in Appendix.~\ref{ref:conv2lstm}.
The four sub-figures in Fig.~\ref{fig:test} demonstrate the comparison between the prediction from the trained Conv2LSTM and ground-truth for testing data, they are the transfer rate $\epsilon$, the infectious rate $\lambda$, the diagnosed rate $ \sigma$ and undiagnosed population $U$ respectively. The well-trained machine extracts parameters from the RKI infection map of March 2020 as $\lambda' = 0.26, \sigma= 0.09 , U_0 = 1352, \epsilon = 0.068$. The first and last parameters are numerically matched with the fitting results, and the diagnosed rate and initial undiagnosed population maintain the consistency with fitting parameters in the relationship $\sigma U_0$ (see Appendix.~\ref{ref:para}).

\subsection{Machine Learning-assisted Risk Prediction}
To assess the COVID-19 risk in realistic situation timely, we deploy the well-trained neural network into the current cases with transfer learning method. Transfer learning is a machine learning method where a model developed for a task is generally reused as the starting point for a model on a second task. In our case, transferring the pre-trained models into the on-going second-wave COVID-19 epidemics in Germany, we extract the dynamical parameters of the spatio-temporal pandemics from the 30-day RKI infection maps with beginning from September 8, 2020, which yields the dynamical parameters as $\lambda' = 0.21, \sigma= 0.167, U_0 = 7426, \epsilon = 0.09$.

\begin{figure}[ht]
\centering
\sidesubfloat[]{\includegraphics[width=3.5cm]{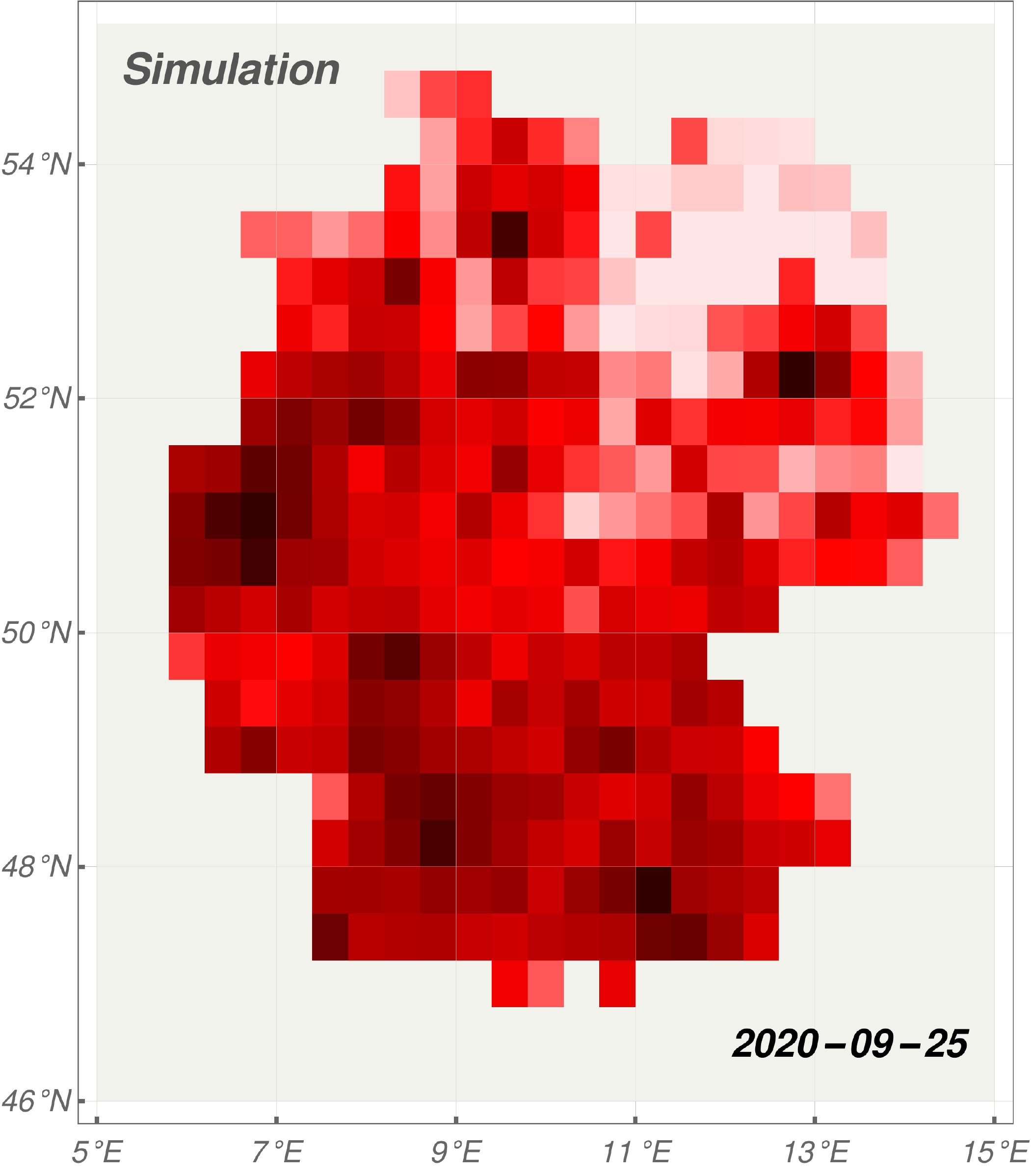}
\includegraphics[width=3.5cm]{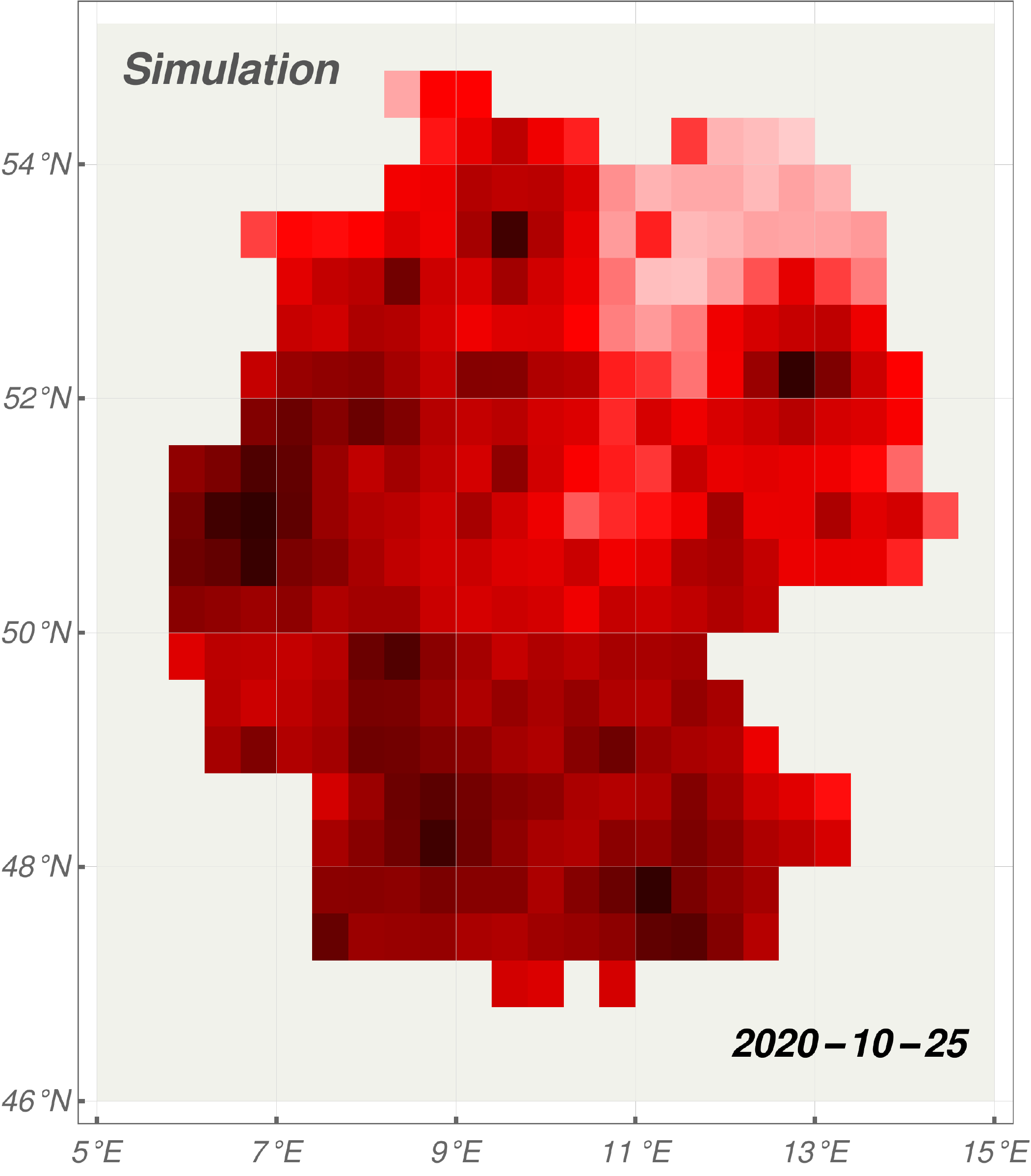}
\includegraphics[width=3.5cm]{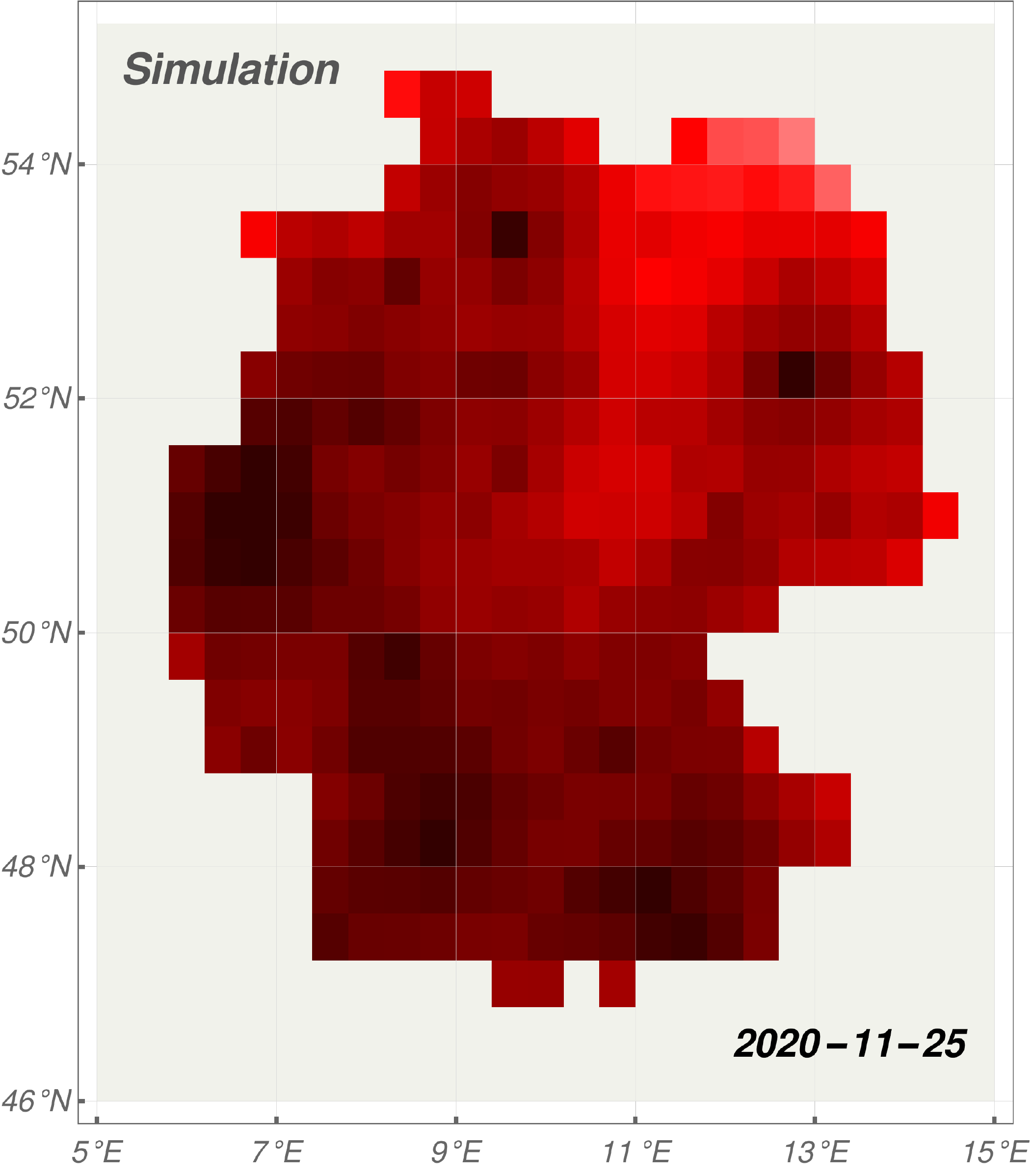}\label{fig:simu}
\hspace{0.25 cm}
\includegraphics[scale = 0.25]{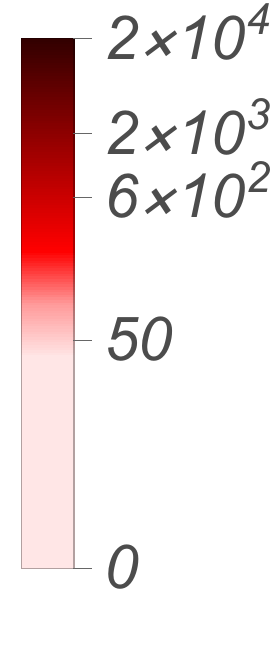}
}
\vspace{0.2 in}
\sidesubfloat[]{\includegraphics[width=3.5cm]{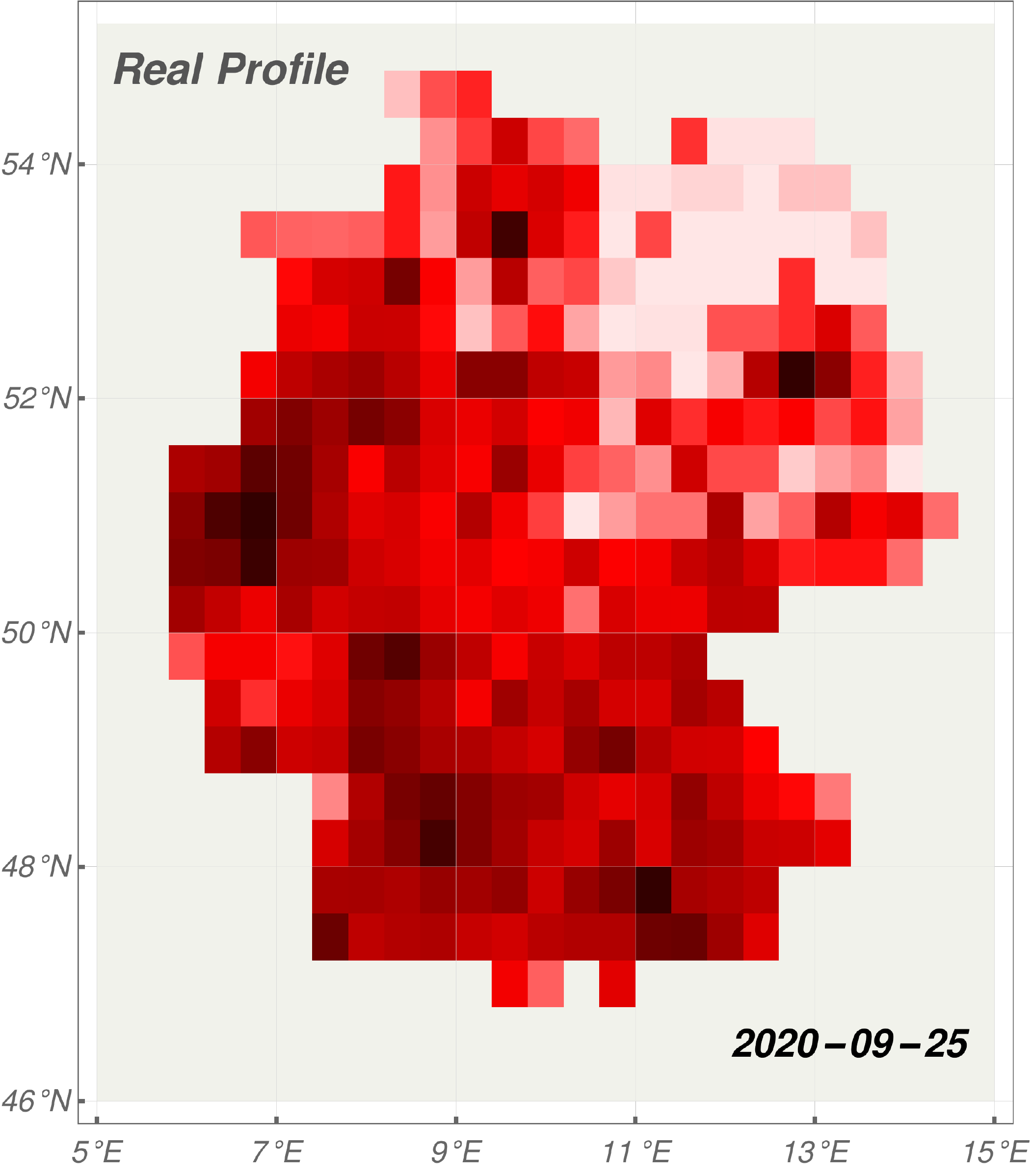}
\includegraphics[width=3.5cm]{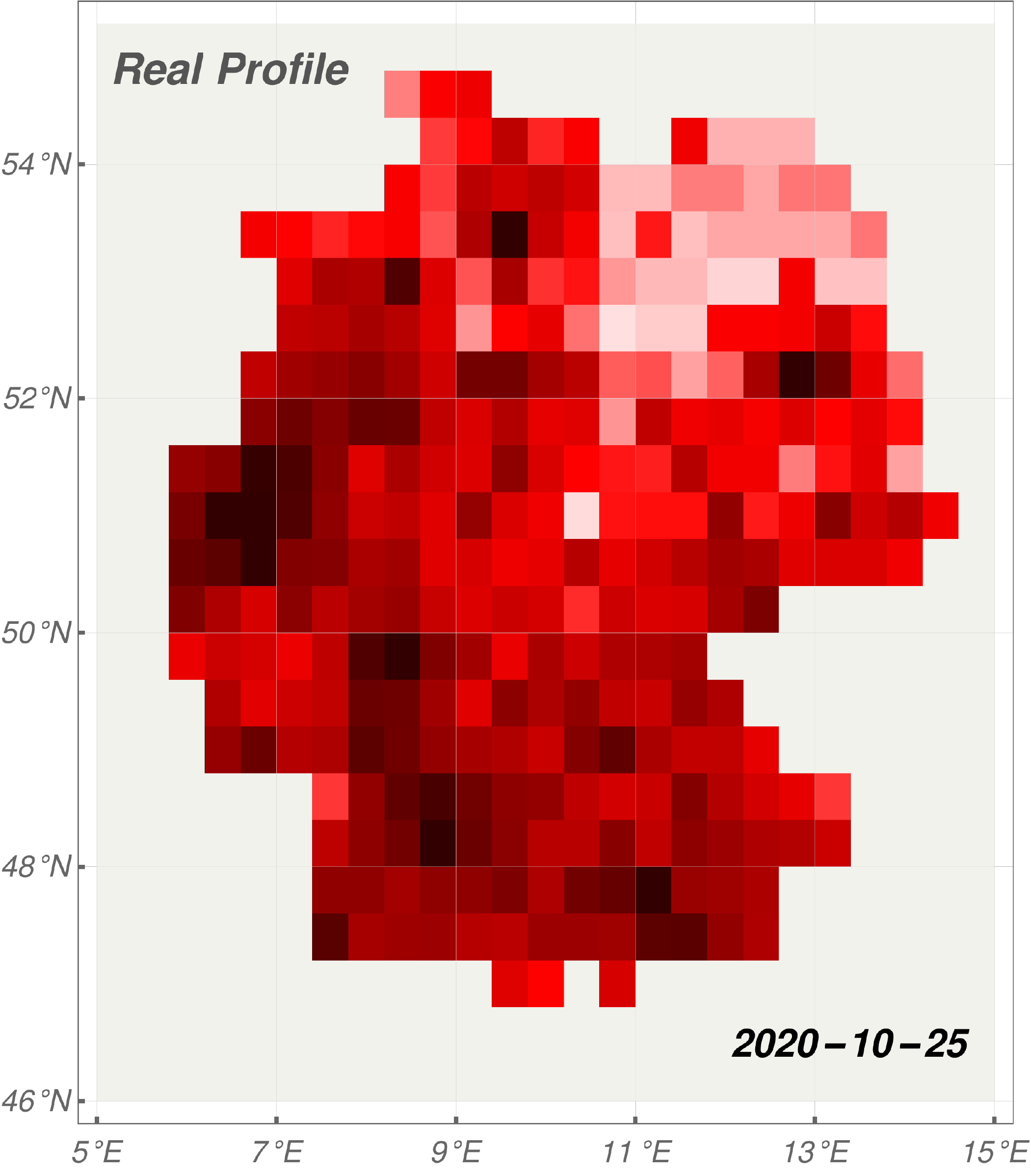}
\includegraphics[width=3.5cm]{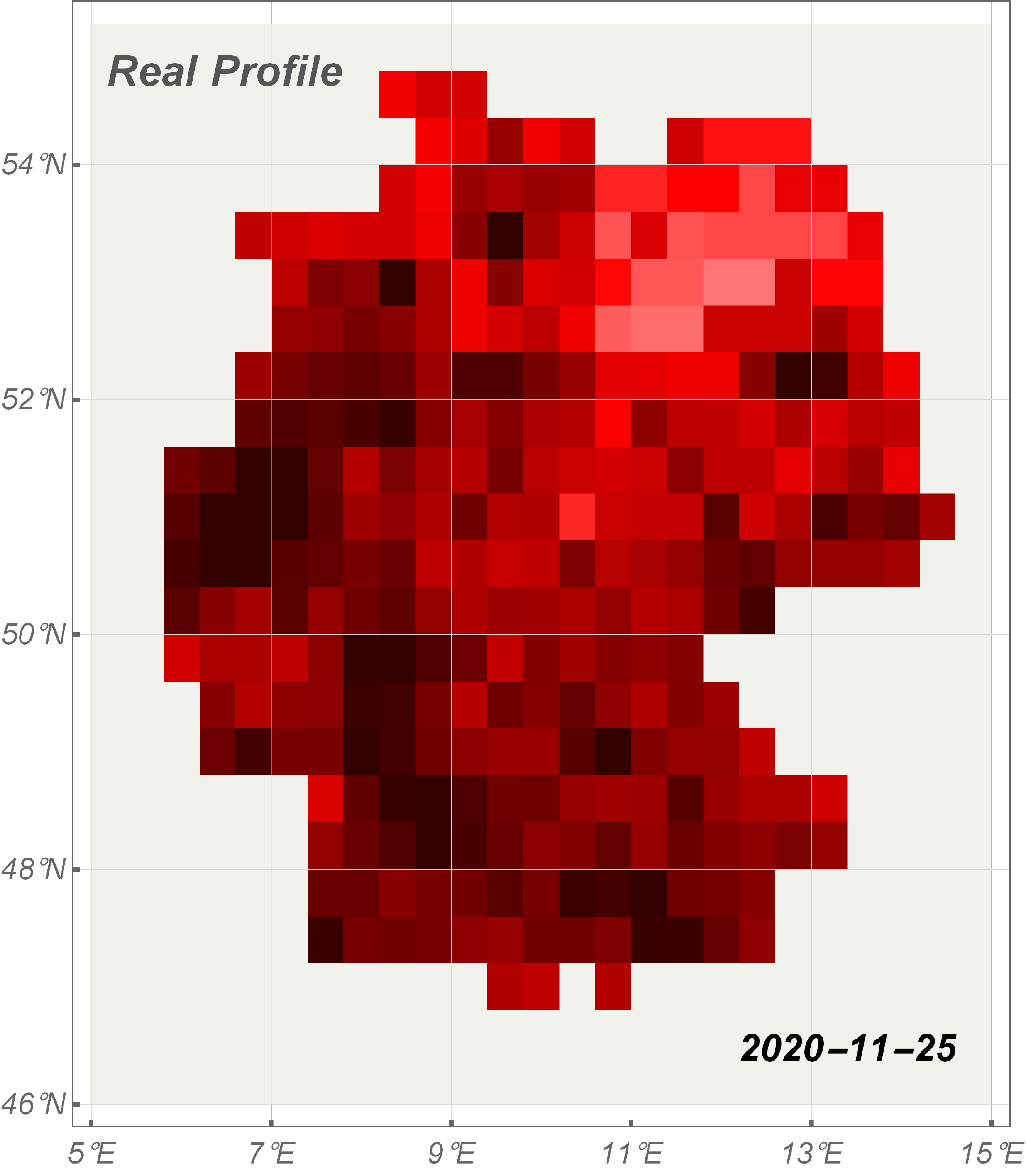}\label{fig:real}
\hspace{0.25 cm}
\includegraphics[scale = 0.25]{maps/vLegend2.pdf}
}
\caption{\textbf{Comparison of simulation and RKI infection maps from September to November 2020 in Germany.} The color bar is set into an uniform red color scale, the darkest color is 20000 cases with a logarithmic re-scale. \textbf{(a)} As of the end of November, the cumulative infection cases are presented in dynamical simulation with parameters learned by machine learning at 2020-09-25, 2020-10-25 and 2020-11-25 respectively. \textbf{(b)} As of the end of November, the cumulative infection cases are presented in real profiles at 2020-09-25, 2020-10-25 and 2020-11-25 respectively. The top 5 biggest cities in Germany, Berlin (13°23E, 52°31N), Hamburg (10°00E, 53°33N), M\"unchen (11°35E, 48°03N), K\"oln (6°58E, 50°56N) and Frankfurt (8°41E, 50°07N) shows a worse situation than the other sites.}
\label{fig:simureal}
\end{figure}

Fig.~\ref{fig:simureal} describes the homogeneity between simulation and RKI infection maps from September to November 2020 in Germany. The former cumulative infection maps shown in Fig.~\ref{fig:simu} are generated from the SUIR model with the machine learning parameters during 3 months. The latter maps shown in Fig.~\ref{fig:real} are projected from the RKI data, in which we present the cumulative infection cases in real profiles until the end of November. The slight mismatching on the map of November is due to the actual soft lock-down policy has been executed from 2 November. The functional regulations will contain the pandemic moderately, which will be examined in the following section under different policies.

\subsection{Public Policies Evaluation}

\begin{table}[htbp!]
\small
\centering
\caption{\textbf{Public Policies for Different Spots.} This table lists specific rules which lead to three scenarios: a) unlocked, b) soft-locked, c) locked. The last one is the well-know lock-down policy which shows most powerful restriction to the mobility of human.}
\label{tab:policies}
\begin{tabular}{@{}cccc@{}}
\toprule
                     & \textbf{Unlocked} & \textbf{Soft-locked}               & \textbf{Locked} \\ \midrule
\textbf{Office}      & Office work in schedule      & Home office for large group        & Home office     \\
\textbf{Campus}      & Open              & Close                               & Close           \\
\textbf{School} & Open              & Close     & Close           \\
\textbf{Supermarket} & Open              & Open with distance constrains      & Close           \\
\textbf{Restaurant} & Customer number constrain              & Only Take-out      & Close           \\
\textbf{Travel}      & Free movement     & Long distance travel is prohibited & Home isolation  \\ \bottomrule
\end{tabular}
\end{table}

Three different degrees of restriction rules, named as the unlocked, soft-locked and locked strategies in public policy, are evaluated in our COVID-19 evaluation framework. The examples of the corresponding policies are listed in Table.~\ref{tab:policies}, in which we list some representative government policies in different spots. It is of crucial concern for both governments and residents who are enduring the social-economical pressure in pandemic~\cite{chande:2020realtime,bavel:2020using,hsiang:2020effect,wong:2020modeling}. The hot spots include but are not limited to office, campus, school, restaurant and cross-county travel. It should be mentioned that although we introduce the heterogeneous rules in different spots the comprehensive influences at county-level are averaged, which means the effect of policies are coarse graining to the county-level. Reasonably, it could be improved if we collect more precise data below county-level, e.g., individual or community-level~\cite{chang:2020mobility}.

\begin{figure}[ht]
\centering
\sidesubfloat[]{\includegraphics[width=4cm]{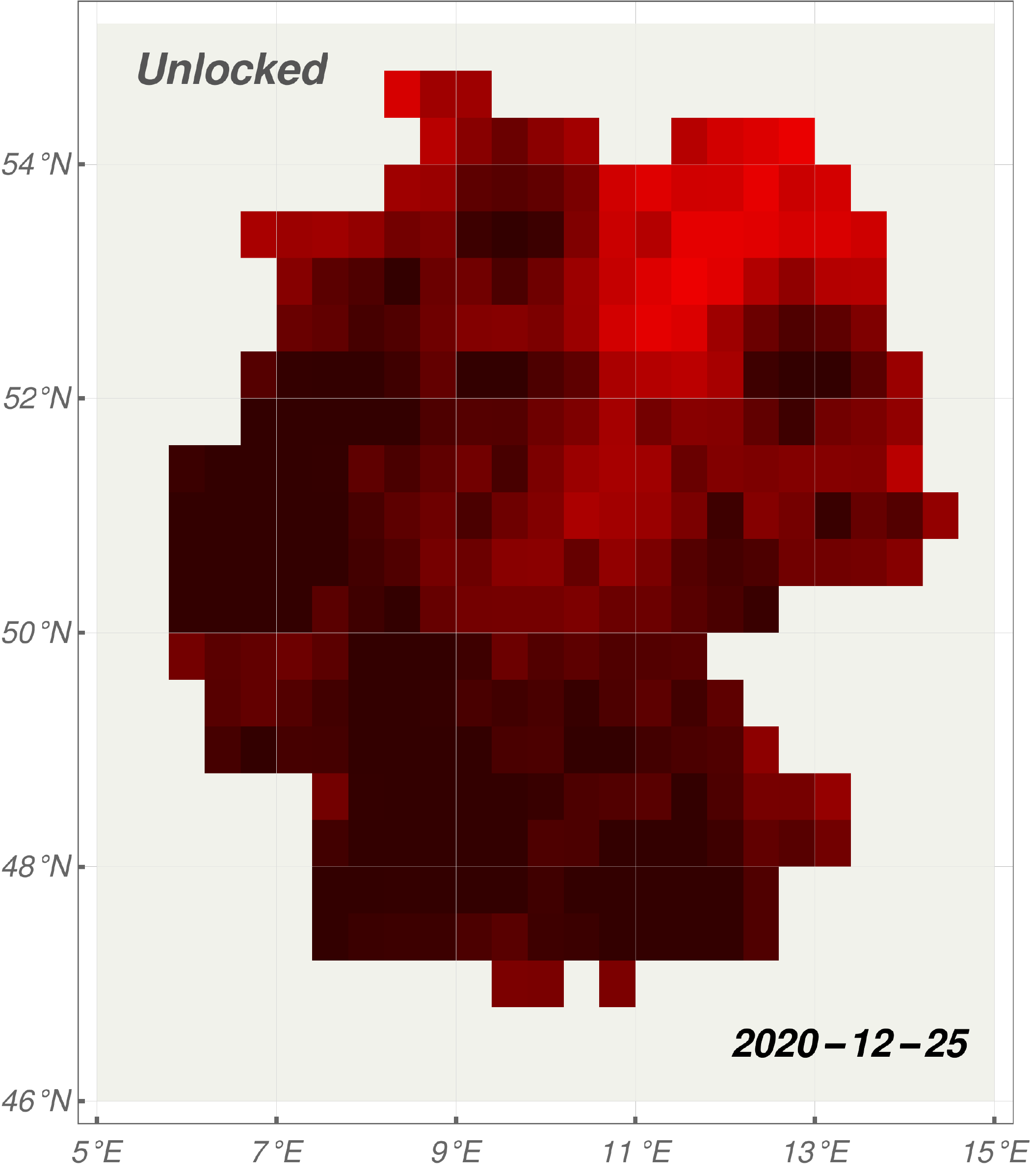}\label{fig:unlock}}
\sidesubfloat[]{\includegraphics[width=4cm]{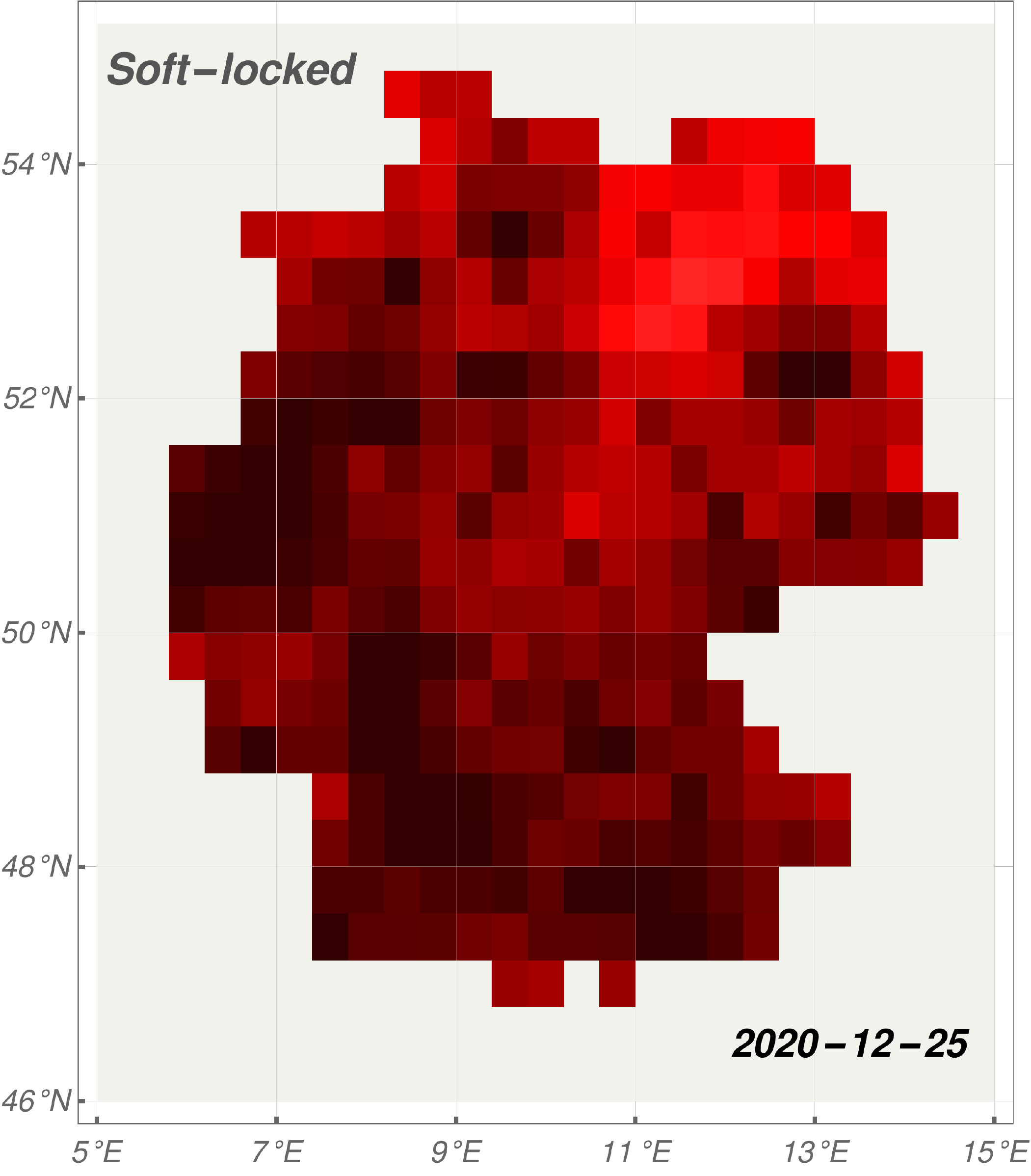}\label{fig:softlock}}
\sidesubfloat[]{\includegraphics[width=4cm]{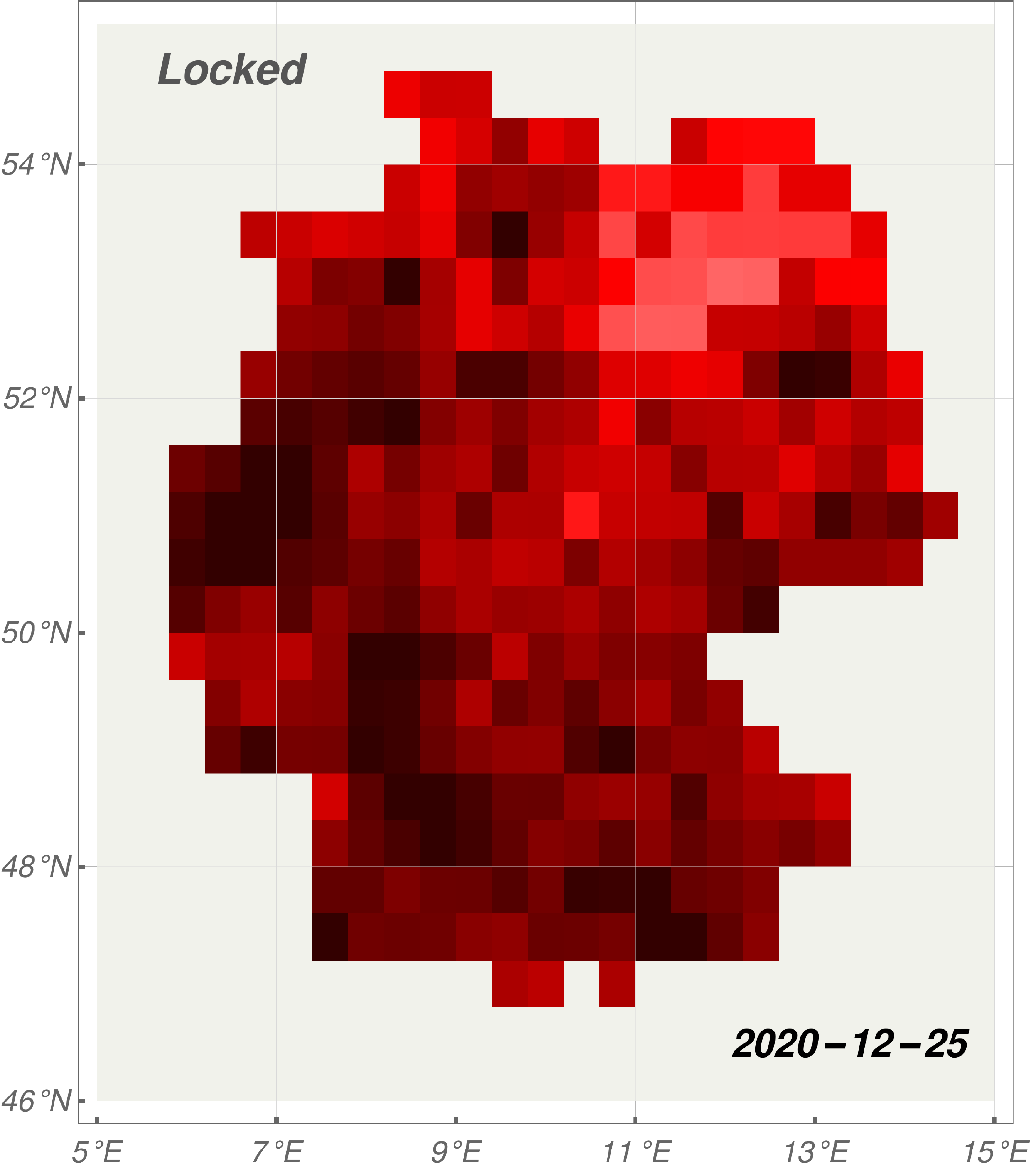}\label{fig:locked}
\hspace{0.25 cm}
\includegraphics[scale = 0.25]{maps/vLegend2.pdf}
}
\caption{\textbf{Predicted cumulative infection maps at Christmas under three different public policies.}  The color bar is set into an uniform red color scale, the darkest color is 20000 cases with a logarithmic re-scale. The dynamical parameters are extracted from RKI infection maps by the well-trained machine. \textbf{(a)} The prediction is based on the existing public policies. \textbf{(b)} The soft lockdown rules contains the relative loose restrictions to some hot spots, thus the diffusion of the epidemic is alleviated partially. \textbf{(c)}Locked: These strictest rules prohibit any form of mobility, which yield the least infections.}
\label{fig:policy}
\end{figure}
The differences among them in simulation are the transfer rate and susceptible population: the most strict locked rules has transfer rate $\epsilon=0$, the soft-locked with half transfer $\epsilon=0.045$, and the rule with no additional restriction has $\epsilon=0.09$ as default; meanwhile, the containment policies could result in a consequence of infectious population decreasing that effectively deplete the susceptible population~\cite{maier:2020effective}. It leads to the reduction of the susceptible population, reaching to $1/2, 1/8$ for the locked and soft-locked rules respectively, in which the $1/8$ is from the rough estimation to the necessary out time. These three strategies are implemented from the same time point 25 November 2020, and their 30-day simulation results are shown in Fig.~\ref{fig:policy}. In our prediction assisted by the transfer learning method, 2.3 million people will be infected before Christmas in the worst situation, where daily increase is approximately 80 thousand (estimate with mortality rate $\sim1.5\%$~\cite{stafford:2020covid19}). Compared with the prediction of the Institute for Health Metrics and Evaluation (IHME)\footnote{http://www.healthdata.org/covid}, they achieve the consistency in order of magnitude, which means the total mortality will reach 34.5 thousand approximately. With regard to the soft-locked policy, the infection number reduces to 1.4 million and the corresponding daily increasing cases are 20 thousand. In the most strict rules, although it restricts the mobility, the number of infection cases will be 1.07 million and the daily increase is 4 thousand.

\section{Discussion}
In this paper, we construct a multi-level spatio-temporal epidemiological model that combines a spatial cellular Automaton  (CA) with a temporal hybrid Susceptible-Undiagnosed-Infected-Removed (SUIR) model which contains the human mobility across the counties. This new toolbox enables the projection of the state-county-level COVID-19 prevalence over 412 Landkreise in Germany, including t-day-ahead risk forecast and the risk evaluation related to a public policy. The modular design in our survey can reach a more accurate estimation of the COVID-19 risk with replacing the modules by the other elaborate spatial or temporal models conveniently. With the help of machine learning methods, we extract the dynamical parameters directly from the infection maps, which are topologically equivalent to the real geographic maps. After training the Conv2LSTM neural networks composed of time distributed CNNs and LSTM on data set generated from CA-SUIR model, we transfer the well-trained neural networks into the data set collected from the open database of RKI. 

The CA-SUIR model reproduces the evolution of the infection maps in March and September acknowledged as the first- and second-wave of COVID-19 pandemic. Even though the prediction focuses on the infectious dynamics in the current paper, it is conveniently feasible to derive recovery and mortality rate from the model under the guarantee of accurate data sources confronted. The simulations perform properly to describe and predict the data in the initial 30-day phase, while it shows a tendency towards a faster smoothing-out of the pronounced local fine structure and persistent hot spots, as compared to the data. It could be understood that the observed infection cases concentrated in the big cities, such as Berlin, Hamburg, M\"unchen, K\"oln and Frankfurt than the other sites in the simulation, which reflects high population density enhances the diffusion of the COVID-19. The prediction could be evidently improved by introducing more abundant interactions across and within the counties. As some practical improvements, e.g. replacing the geographic lattice to traffic networks~\cite{estrada:2020covid19,chang:2020mobility,topirceanu:2020centralized,valba:2020selfisolation,ye:2020modeling}, introducing the higher resolution lattices beyond the county-level which also means higher-order interactions among the residents~\cite{rader:2020crowding,schneckenreither:2008modelling,chang:2020mobility,charoenwong:2020social}, can be implemented and devoted in further future works. With the help of the deep learning, the machine learns the dynamical parameters matched with the data of March, which is transferred into the prediction to second-wave sequentially. Starting in September, the simulation maps are remarkably consistent with the data. Regarding the slight uncertainty in extracting parameters for the dynamics in September, which might has root in the difference of development trend between the first- and second-wave, it could be avoided by training the neural networks on more heterogeneous data set. We will present the further improvement in our future papers.

Compared to the prevailing COVID-19 risk prediction models, the machine learning-assisted CA-SUIR model here indicates that knowledge of risk evaluation in the county-level can be displayed directly to governments and residents. The prediction to the effects of different public health policies suggests that the transmission modes of coronavirus could be shaped by the efficient non-pharmaceutical interventions. Besides, the 7-day windows could give a robust prediction in long-term, which is advantageous to timely assess the current situation. The other valuable application of our evaluation system is to improve allocation of medical resources which are routinely unequal in different counties, some of them are suffering from the epidemic that far exceeds their medical capacity. It is of essential significance to implement the intervenable risk evaluation model for decision-making on restarting economics and public health policies in the COVID-19 pandemic.

\section{Acknowledgments}
The authors thank Esteban Vargas and Maria Babarossa for useful discussions and comments. The work on this research is supported by the BMBF through the ErUM-Data funding and the Samson AG AI grant (L.W, K.Z), by the National Natural Science Foundation of China, Grant No. 11875002(T.X, Y.J.), by the Zhuobai Program of Beihang University(Y.J.). The authors also thank the donation of NVIDIA GPUs from NVIDIA Corporation.

\appendix

\section{Infection Maps}\label{ref:top}

\begin{figure}[htbp!]
	\centering
	\includegraphics[width=9cm]{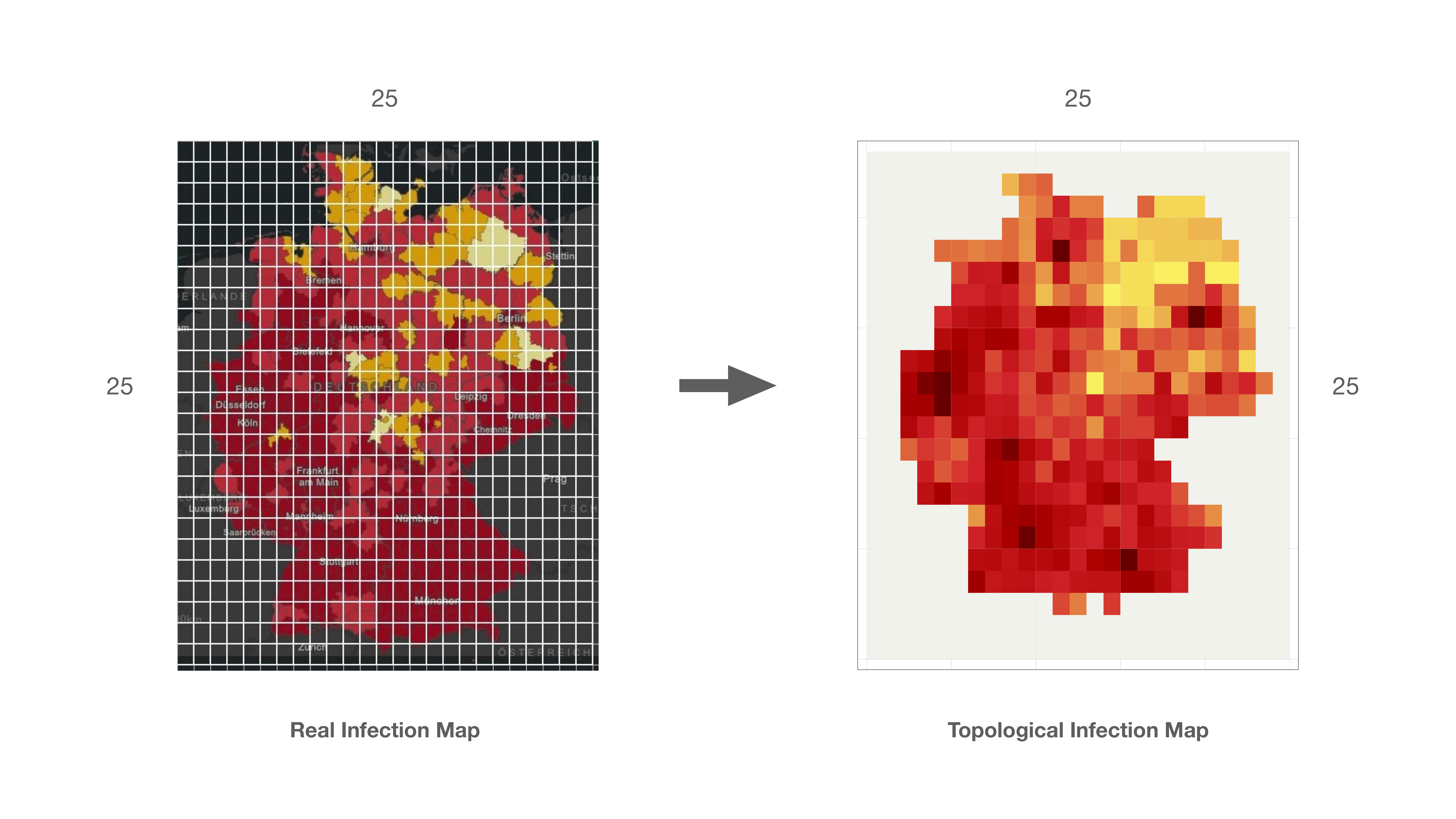}
	\caption{\textbf{Mapping from real infection map of Germany onto topological equivalent lattice.}  The left map is divided into $25\times25$ sites as the size of the right part. The geometric shape of the suqare lattice is irrelevant to the real geographic topography. The site on lattice depict a typical county-level unit of the real map.}
	\label{fig:map}
\end{figure}

In order to simulate the population evolution on lattice, the topological mapping is adopted in our CA computation. We firstly embed the map in a rectangle area whose length and width are chosen as the maximum value of the Germany map, as in Fig. \ref{fig:map}(left). Then the area is segmented into uniform sites of $L \times L$. And each site is labelled by its row and column numbers as $B_{m n}$, where $1\leq m, n\leq L$. If a county occupies $N$ sites, all kinds of the population, such as susceptible, undiagnosed, infected and removed, are assigned equally into each site. And if there is more than one county or county part on a site, the populations of the site is set as the sum of all these counties or county parts.  Because the geographic distance is not important in our simulation, which focus on the population distributions, we set each site as a square when plotting the distribution in Fig. \ref{fig:map}(right). In the current simulation, we confine all the population in Germany and thus set the out-of-Germany transfer possibility at the border as zero.

\begin{figure}[htbp!]
\centering
\sidesubfloat[]{\includegraphics[width=3cm]{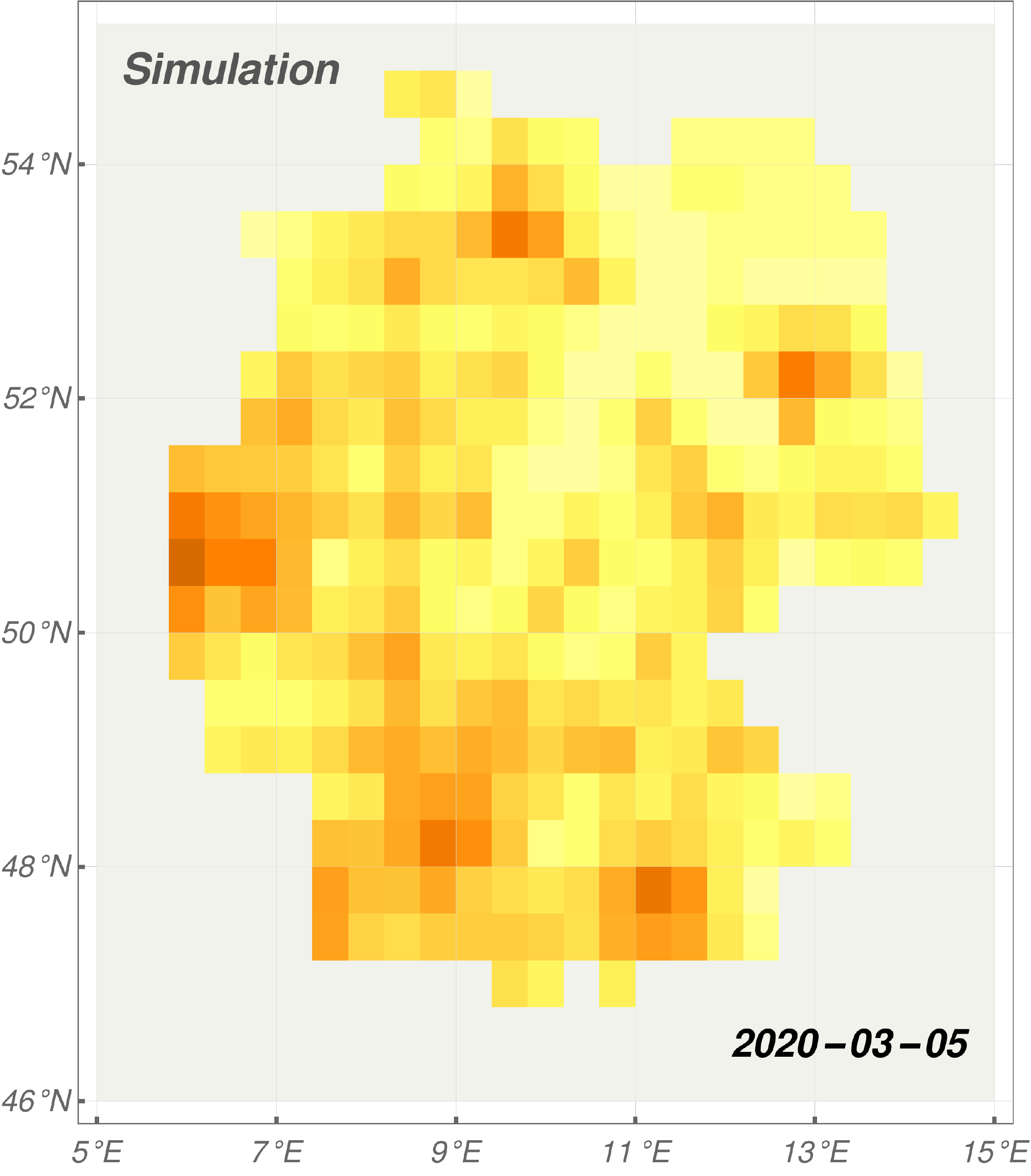}
\includegraphics[width=3cm]{maps/Sim_03_15.pdf}
\includegraphics[width=3cm]{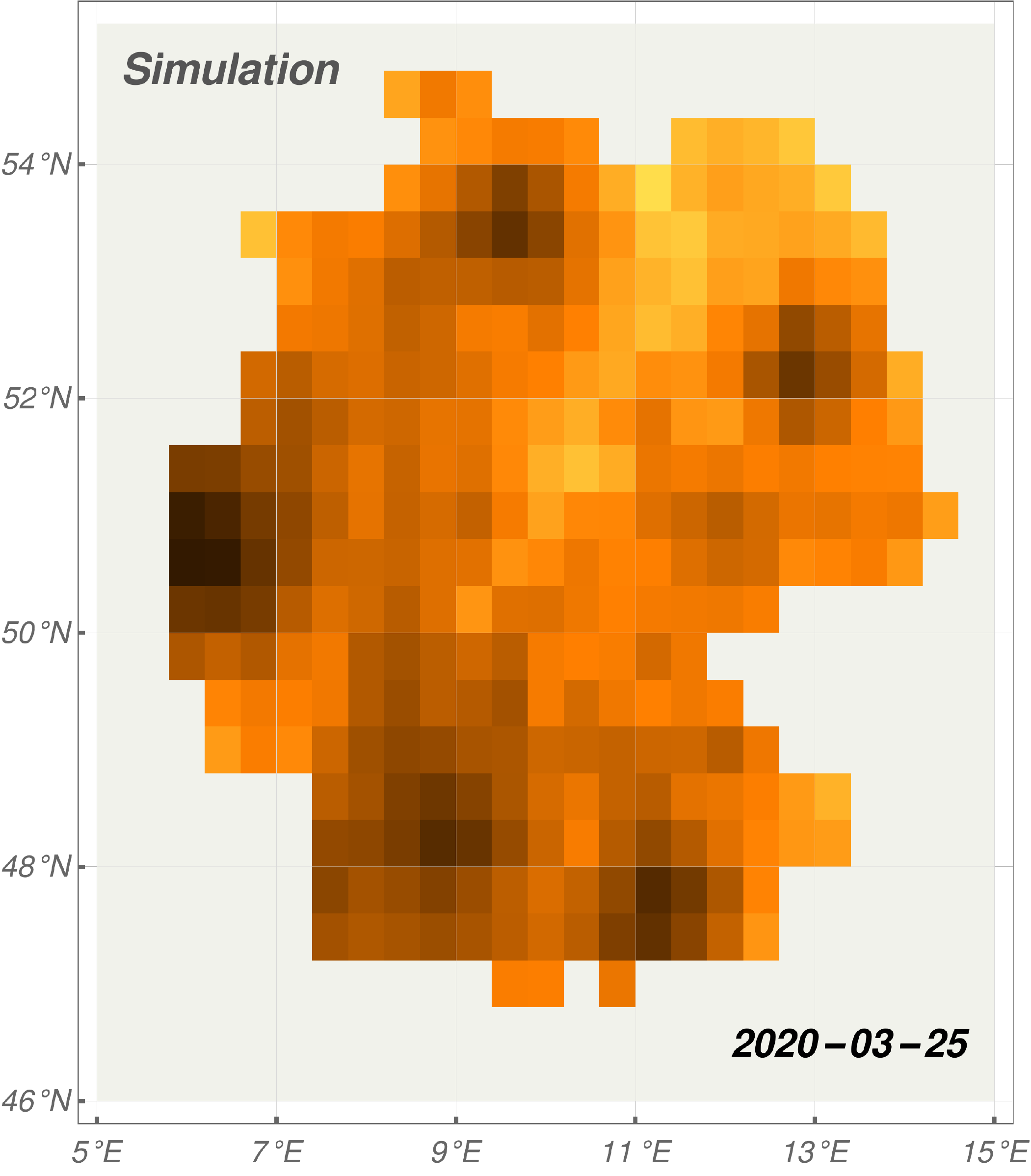}
\hspace{0.25 cm}
\includegraphics[scale = 0.25]{maps/vLegend1.pdf}
}
\vspace{0.2 in}
\sidesubfloat[]{\includegraphics[width=3cm]{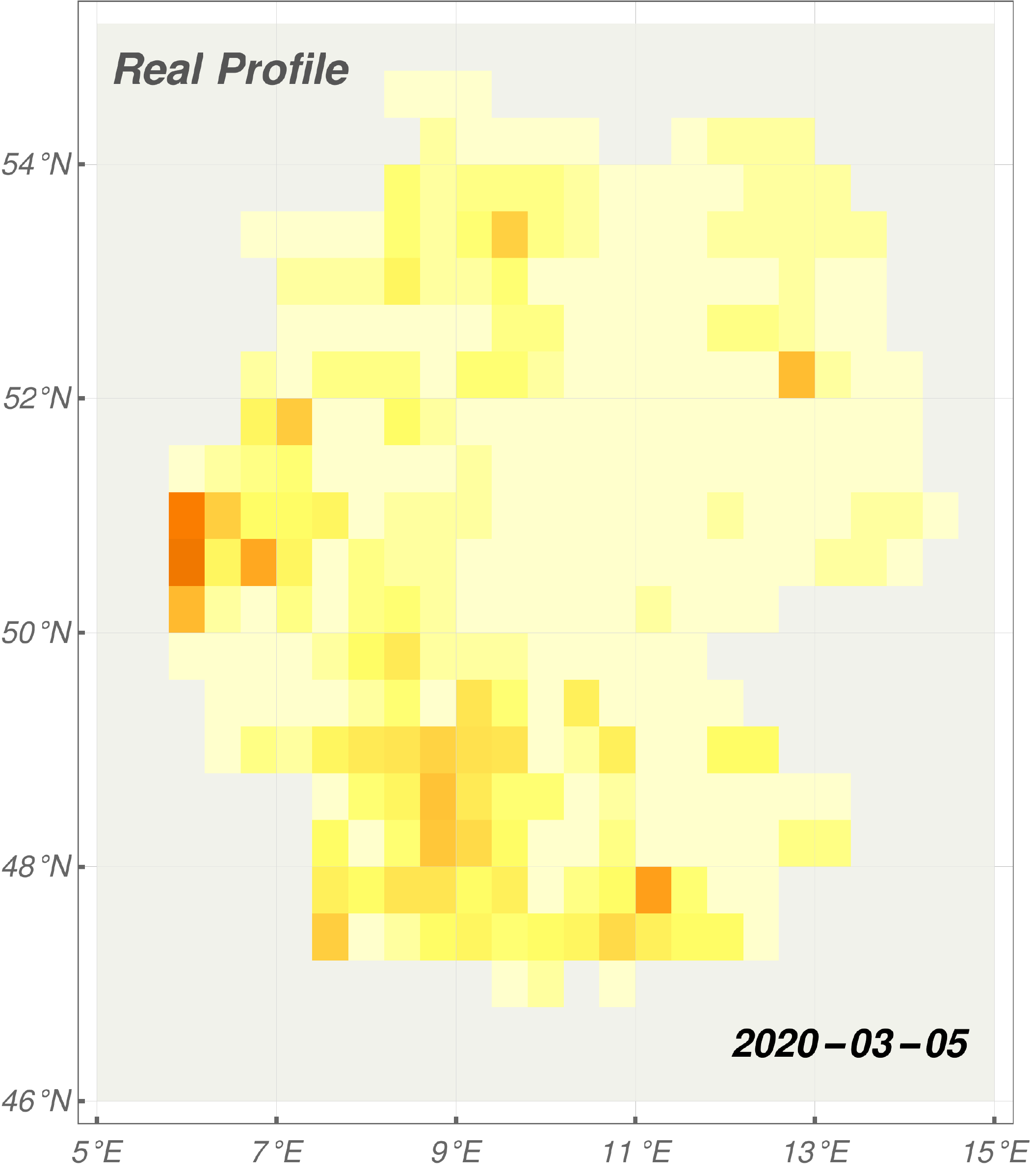}
\includegraphics[width=3cm]{maps/Rel_03_15.pdf}
\includegraphics[width=3cm]{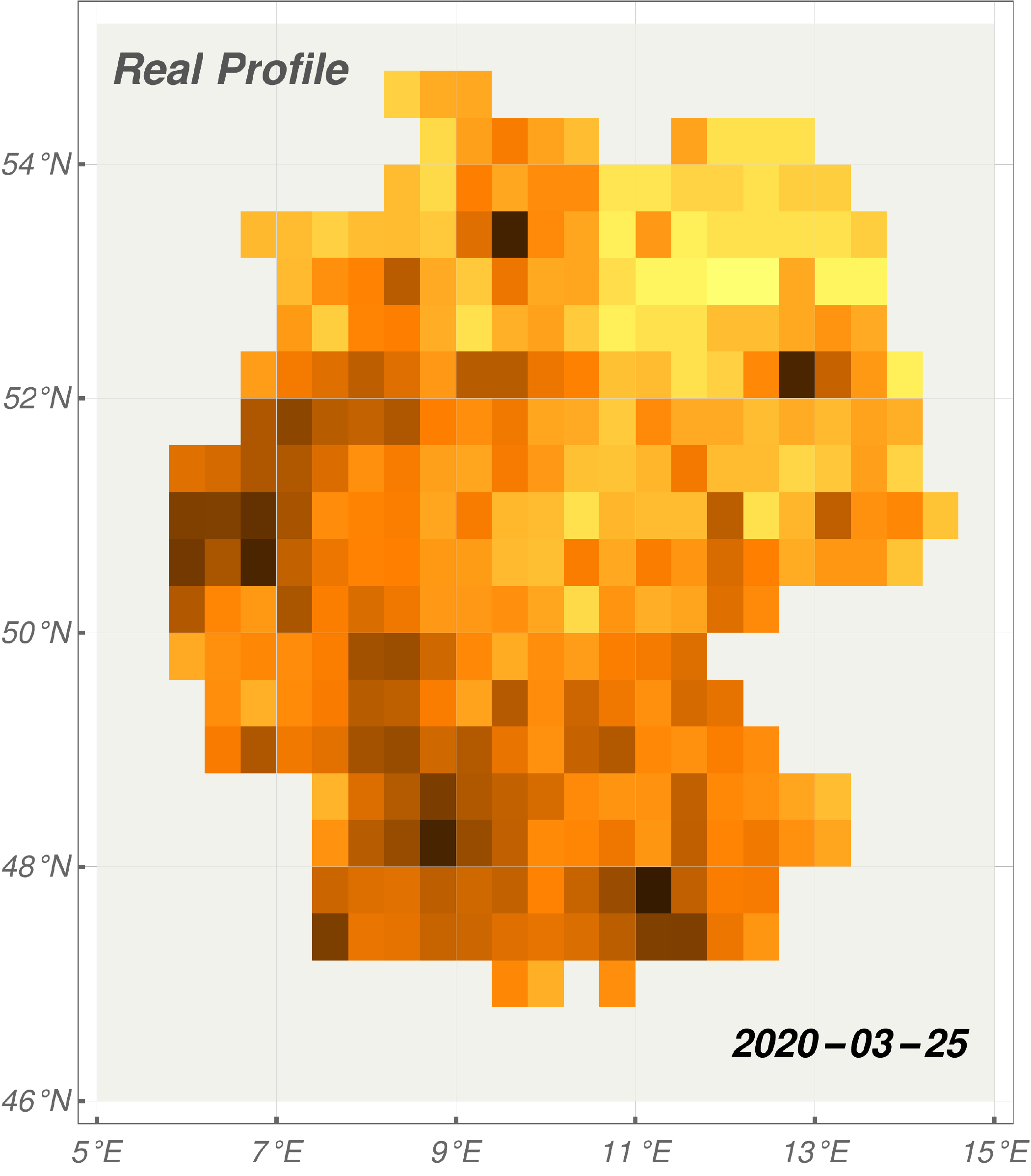}
\hspace{0.25 cm}
\includegraphics[scale = 0.25]{maps/vLegend1.pdf}
}
\caption{\textbf{Comparison of simulated and RKI infection maps during March 2020 in Germany.} The color bar is set in an uniform orange color scale, the darkest color is 2000 cases with a logarithic re-scale. \textbf{(a)} As of the end of March, the cumulative infection cases are presented in dynamical simulation with fitting parameters at 2020-03-05, 2020-03-15 and 2020-03-25 respectively. \textbf{(b)} As of the end of March, the cumulative infection cases are presented in real profiles at 2020-03-05, 2020-03-15 and 2020-03-25 respectively. }
\end{figure}

\section{SUIR model parameters explanation}\label{ref:para}
In this part, we discuss the fitting details based on the mathematical simplification to the SUIR model. In the large susceptible population limit and ignoring the recovery and mortality cases for the uncertainty at early stage, there are only two equations are relevant, 
\begin{align}
\frac{d U}{d t} &=(\lambda  -\sigma - \frac{\sigma'}{\text{days}})U  \\
\frac{d (I+R)}{d t} &=\sigma U
\end{align}

The solutions are clearly derived as,
\begin{align}
U(t) &=U_0 Exp[(\lambda  -\sigma - \frac{\sigma'}{\text{days}})t]  \\
I+R &=I_0+R_0-\frac{\sigma U_0}{\lambda  -\sigma - \frac{\sigma'}{\text{days}}}+\frac{\sigma U(t)}{\lambda  -\sigma - \frac{\sigma'}{\text{days}}}
\end{align}

The $I+R$ fitting is easy now with $I+R=a_* +b_* Exp[c_* t]$. This means we can not obtain all the independent parameters in the solution, but the following redefinition could be helpful,
\begin{align}
a_* &=I_0+R_0-\frac{\sigma U_0}{\lambda  -\sigma - \frac{\sigma'}{\text{days}}}\\
b_* &=\frac{\sigma U_0}{\lambda  -\sigma - \frac{\sigma'}{\text{days}}}\\
c_* &=\lambda  -\sigma - \frac{\sigma'}{\text{days}}
\end{align}
or equivalently only the following combinations could be obtained
\begin{align}
&I_0+R_0 = a_*+b_*\\
&\sigma U_0 =b_* c_*\\
&\lambda  -\sigma - \frac{\sigma'}{\text{days}} =c_*
\end{align}
As for the product $\sigma U_0$, which means it is difficult to determine the $\sigma$ and $U_0$ individually. It will introduce the inevitable uncertainty to the two parameters, while the confidence to the multiplicative variable could be reserved.

\newpage

\section{Conv2LSTM Neural Networks}\label{ref:conv2lstm}
\begin{figure}[htbp!]
	\centering
	\includegraphics[width=13cm]{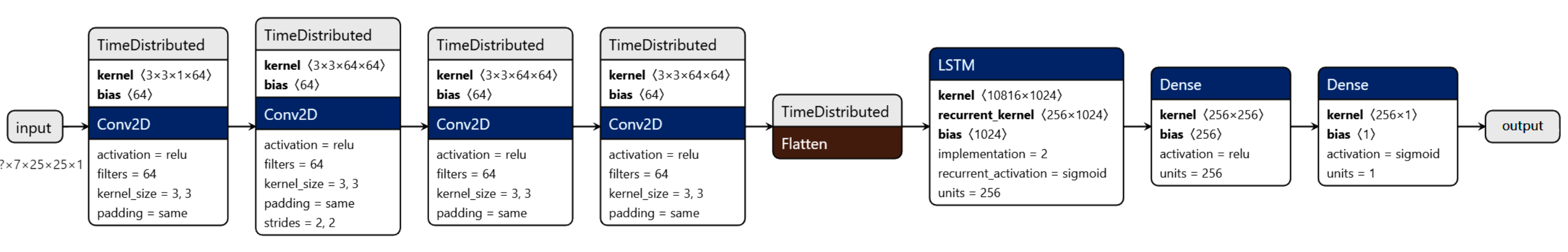}
	\caption{\textbf{The Conv2LSTM neural networks.} The nuts-and-bolts of the neural network are: Convolution layers within TimeDistributed wrapper, from Convolution 2D (64, kernel size=3x3, `ReLU', `same' padding) to Convolution 2D (64, kernel size=3x3, strides=2, `ReLU', `same' padding) to Convolution 2D(64, kernel size=3x3, `ReLU', `same' padding) to Convolution 2D(64, kernel size=3x3, `ReLU', `same' padding) and Flatten; for the LSTM part shown in Fig.~\ref{fig:nn}, 256 LSTM cells with `sequence=False' and activation=`tanh' are used; dense layers contains 256 neurons with `ReLU' active function, before the output is the `sigmoid' function to fit the normalized targets.}
	\label{fig:nn}
\end{figure}
\begin{figure}[htbp!]
	\centering
	\includegraphics[width=9cm]{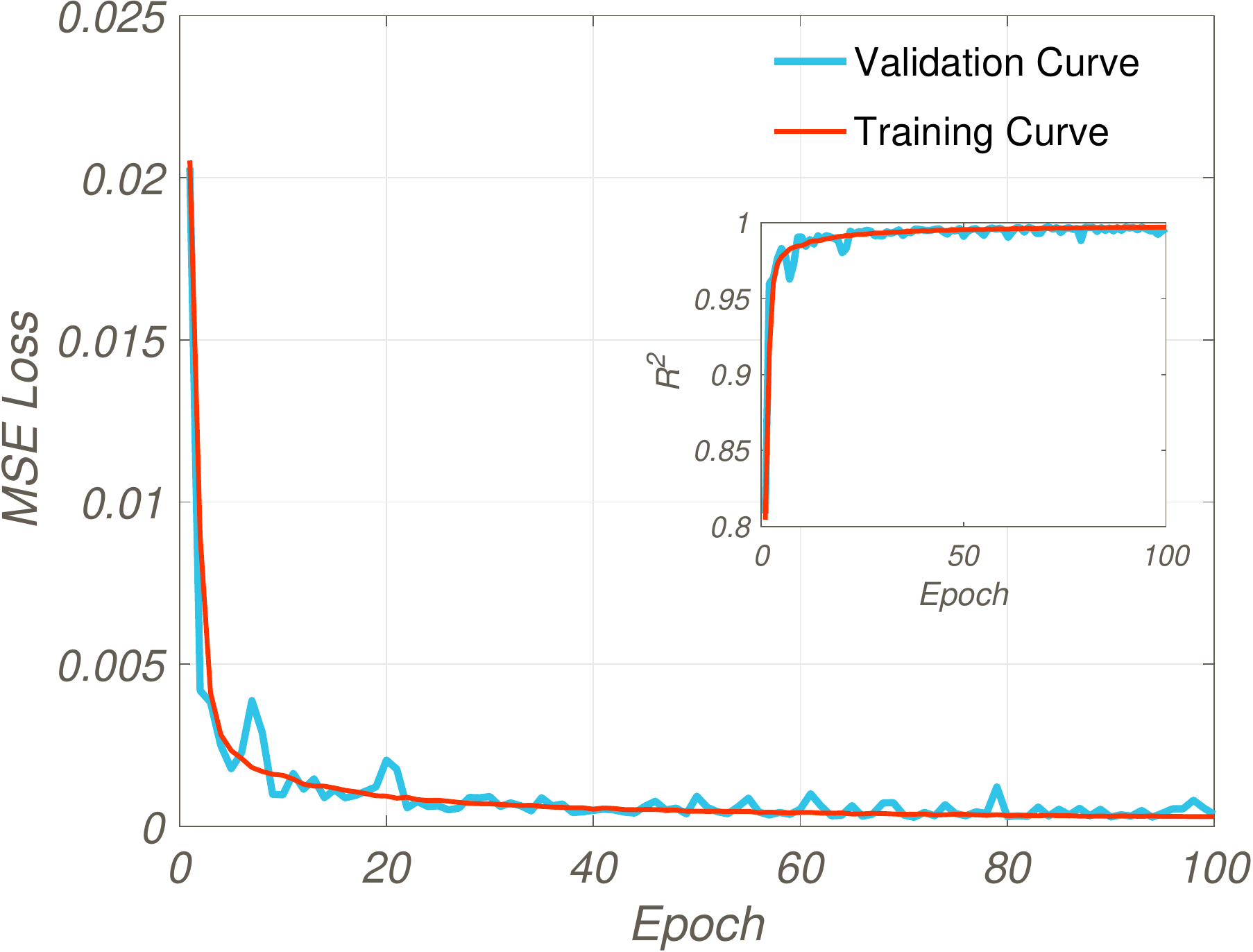}
	\caption{\textbf{Learning curve of the Conv2LSTM neural networks.} The red line is the learning curve, in which the Mmean-Square Error (MSE) loss is reducing with the epoch increasing. The blue line is located in validation data set, which shows the similar results in learning data set. The increasing of the R square with training is shown in subfigure, which indicates the correlation between the prediction from neural networks and truth tends to be close.}
	\label{fig:learncurve}
\end{figure}

\newpage

\bibliographystyle{elsarticle-num-names}
\bibliography{COVID}

\end{document}